\documentclass[lettersize,journal]{IEEEtran}
\IEEEoverridecommandlockouts

\newif\ifarxiv\arxivfalse

\usepackage{cite}

\ifCLASSINFOpdf
\else
\fi

\usepackage{amsthm}

\usepackage{amsmath,amssymb,mathtools,dsfont,bm}

\usepackage{color}

\usepackage{algorithm,algorithmic}

\DeclarePairedDelimiter{\norm}{\lVert}{\rVert}

\usepackage{subcaption}
\usepackage{enumerate}
\usepackage{amssymb}
\usepackage{mathtools}
\usepackage{orcidlink}

\usepackage{xcolor}
\usepackage{colortbl}

\title{Uncertainty-Aware Federated Learning for Joint Communication and Computation in XR Systems}

\author{Robert-Jeron Reifert\,\orcidlink{0000-0003-3922-8996},~\IEEEmembership{Graduate Student Member,~IEEE,}
	Hayssam Dahrouj\,\orcidlink{0000-0002-0737-6372},~\IEEEmembership{Senior Member,~IEEE,}\\
	Alaa Alameer Ahmad\,\orcidlink{0000-0002-0764-5560},
	Haris Gacanin\,\orcidlink{0000-0003-3168-8883},~\IEEEmembership{Fellow,~IEEE,} and
	Aydin Sezgin\,\orcidlink{0000-0003-3511-2662},~\IEEEmembership{Senior Member,~IEEE}
\thanks{
	This paper has been accepted for publication in IEEE Transactions on Mobile Computing.\newline
	\copyright 2026 IEEE.  Personal use of this material is permitted.  Permission from IEEE must be obtained for all other uses, in any current or future media, including reprinting/republishing this material for advertising or promotional purposes, creating new collective works, for resale or redistribution to servers or lists, or reuse of any copyrighted component of this work in other works.\\
	Part of this paper was presented at the 19th International Symposium on Wireless Communication Systems, July 2024 \cite{10639072}.\newline
	This work was partially funded by the German Federal Ministry of Research, Technology and Space (BMFTR) in the course of the 6GEM+ Transfer Hub under grant 16KIS2411.\\
	Robert-Jeron Reifert and Aydin Sezgin are with Digital Communication Systems, Ruhr University Bochum, Bochum, Germany. (\{robert-.reifert,aydin.sezgin\}@rub.de). 
	Hayssam Dahrouj is with the Department of Electrical Engineering, University of Sharjah, Sharjah, United Arab Emirates. (hayssam.dahrouj@gmail.com). 
	Alaa Alameer Ahmad is with Cariad SE, Wolfsburg, Germany. (alaa.alameer85@gmail.com). 
	Haris Gacanin is with the Chair for Distributed Signal Processing, RWTH Aachen University, Aachen, Germany. (harisg@dsp.rwth-aachen.de).}}

\begin{document}
\maketitle

\begin{abstract}
	The convergence of communication and computation, empowered by artificial intelligence (AI) enhanced capabilities, emerges as a key enabler for realizing the promising functionalities of the sixth-generation of communication systems (6G). To this end, this paper investigates an AI-native, robust network, where multiple base stations (BSs) simultaneously serve several extended reality (XR) devices. The paper then presents an innovative distributed deep learning-based approach to simultaneously manage the transmit and computing powers, and computation allocation, amidst uncertainties in both channel and computing states information. Under such an interference network scenario, the paper proposes a robust approach for minimizing the worst-case delay across the served XR devices, subject to computation and power constraints. The paper leverages a deep neural network (DNN) that can be trained either centrally or in a federated manner to map estimated channels and computation requirements to optimized resource allocations per BS. During training, uncertainty samples, reflecting both channel and computing estimation errors, are injected after the DNN output to account for such joint uncertainties. Our results validate the enhanced robust delay performance and scalability of joint uncertainty injection as compared to uncertainty-agnostic methods and state-of-the-art benchmarks, with federated learning-based solutions achieving near-centralized performance while enhancing efficiency, privacy, and adaptability.
\end{abstract}

%
\IEEEpeerreviewmaketitle

\section{Introduction}
As the number of consumers combining smartphones and extended reality (XR) devices is set to double within five years, with three-quarters seeking XR experiences beyond home settings, the demand for cellular XR is rapidly accelerating \cite{ericsson_xr_report}. Delivering gigabit-per-second data rates, millisecond-level latency, and high computational efficiency, XR is expected to drive the sixth-generation of communication systems (6G) toward major architectural and operational advancements \cite{akyildiz2022wireless}. This evolution requires substantial amounts of remote data connectivity and processing, underscoring the need for an integrated communication and computation platform \cite{10666455}. Additionally, 6G is envisioned to comprise widespread ultra-dense networks with numerous base stations (BSs) and devices \cite{9369324}. Efficient coordination within these networks is essential to prevent communication backbone overload while mitigating large-scale interference in real-time. Consequently, automated joint management of communication and computing resources is critical for future decentralized systems.

In response, the International Telecommunication Union (ITU) steers future networks (6G and beyond) toward native integration of machine learning (ML) and artificial intelligence (AI), enabling AI-native air interfaces and ML-driven resource management \cite{IMT-2030}. Deep learning, in particular, offers efficient and near-optimal wireless resource management \cite{10071987}, but often assumes perfect knowledge of channel conditions and computing requirements \cite{10328648}. In practice, robust resource allocation strategies must be uncertainty-aware, i.e., explicitly account for uncertainties in both communication (e.g., missing or outdated channel state information \cite{6036193}) and computation (e.g., fluctuating processing demands \cite{10007803}). While prior works have addressed either communication or computation uncertainty separately, only a few have considered their joint effects \cite{9615112}. From an AI/ML perspective, such uncertainties further propagate into learning-based decision-making. Since training data is often imperfect, ML-based architectures, as explored in \cite{10071987}, must operate under uncertainty in estimated channel states. However, a systematic approach to jointly handling communication and computation uncertainties in AI-driven optimization remains largely unexplored. To bridge this gap, this paper addresses such issues via a joint uncertainty injection scheme in a multi-BS, interference-limited network scenario, generalizing the single BS scenario considered in \cite{10639072}. 

As 6G networks tend to embrace data processing decentralization, e.g., by pushing intelligence toward the edge, the paper proposed solutions are poised to provide a valuable data-driven, robust framework for managing multi-BS deployments \cite{Journal}. In such decentralized networks, relying on traditional centralized learning would require transmitting raw data to a central entity, which is impractical due to privacy issues, excessive bandwidth consumption, and scalability limitations. Federated learning (FL) addresses these challenges allowing individual BSs to train local models using their own data and share only model updates with a central server \cite{10857305,9413539,elkordy2022much}. This decentralized paradigm significantly reduces communication overhead while enhancing both scalability and security. 
Although FL aggregates model parameters globally, the resulting allocations remain heterogeneous across BSs, since each BS performs inference on distinct local data, e.g., local channel estimations. Thus, FL harmonizes training without enforcing identical allocation policies \cite{kairouz2021advances}. 

Along such lines, this paper develops a robust ML-based approach for joint communication and computation resource allocation in delay-sensitive networks with a particular focus on XR systems, accounting for uncertainties in both channel and computing states. More specifically, the paper goal is to minimize the worst-case delay across all XR devices. The paper then formulates both centralized and FL-based optimization frameworks, incorporating uncertainty injection to enhance robustness under both channel and computation estimation errors. The resulting problems are solved using unsupervised learning by means of adopting customized deep neural networks (DNN) trained either centrally or in a federated manner.\looseness-1\vspace*{-.1cm} 

\subsection{Related Works}

Within the International Mobile Telecommunications-2030 (IMT-2030), XR is recognized as a key driver of 6G, bringing forth challenges such as mobility, blockages, and AI integration \cite{akyildiz2022wireless}. To address mobility and blockages, \cite{10328648} proposes a cooperation scheme where strong devices assist weaker ones to improve communication efficiency. Building on this, \cite{10857305} combines decentralized online learning with caching to reduce the XR delay. Minimizing the XR service latency is explored via cloud-based \cite{9363888} and edge-based offloading \cite{10033423}.
A key challenge introduced by XR, however, is the need for joint management of communication and computation resources, as explored in recent works \cite{10328648,6775376,6036193,10007803,9615112}. 
While \cite{10328648} assumes perfect knowledge of channels and computing requirements, which is unrealistic in practice, robust approaches have addressed these limitations individually. For communication, references \cite{6775376,6036193} tackle imperfect channel state information. In contrast, \cite{10007803} introduces a framework to meet task deadlines under processing uncertainty. While \cite{6775376,6036193,10007803} address either communication or computation uncertainty, \cite{9615112} broadens the scope by adopting robust edge processing of video streams before being streamed to the devices. 

Despite their numerical merits, \cite{10328648,6775376,6036193,10007803,9615112} adopt classical optimization techniques. More recently, the integration of AI/ML with next-generation networks is shown to offer scalable and adaptive alternatives, particularly suited for complex tasks like advanced radio resource management \cite{10255239}. For instance, \cite{8755300} investigates the use of neural networks in wireless settings, while \cite{9475121} applies FL for the joint optimization of communication and computation resources. Given that training data is inherently imperfect, \cite{10071987} proposes an uncertainty injection framework in deep learning to enhance the robustness under channel estimation errors. Yet, the joint impact of communication and computation uncertainties remains largely unexplored, with their coupled effects within the delay objective and shared resource constraints being technically challenging. Based on our conference version \cite{10639072}, we address such gap by jointly considering both channel and processing uncertainty in a multi-BS, interference-limited setting.\looseness-1

In fact, the above-discussed works, including \cite{10639072}, rely on traditional centralized architectures, which come with limiting factors such as front- and backhaul bottlenecks, cloud latency, and privacy concerns. Future networks, however, are poised to spearhead decentralized network architectures, offering advantages such as flexibility, resilience, and lower latency \cite{10077210}. In  particular, a promising approach is FL, which enables efficient knowledge aggregation across devices without requiring continuous interaction or exploration-based updates \cite{9141214}. The work \cite{10636212} provides a survey of recent advances in optimization methods for wireless communications, eventually encompassing distributed learning. FL enhances data privacy by ensuring that original training data remains within its source domain \cite{9413539}, while also enabling secure aggregation \cite{elkordy2022much}. For instance, \cite{8917592} applies FL-based resource allocation for ultra-reliable low-latency communication, with significantly reduced data exchange compared to centralized benchmarks. 

Recently, some works jointly address the challenges of integrated communication and computation under uncertainty using ML/AI-aided approaches. This includes optimizing communication and computation resources using FL for energy efficiency \cite{9475121} and employing quantization to minimize the convergence time \cite{10075480}. Furthermore, \cite{9210812} accounts for packet errors, while \cite{9120649} includes channel uncertainty for joint client scheduling. Lastly, \cite{10857305} investigates personalized FL for cellular VR, integrating online learning and field-of-view caching. While some works (e.g., \cite{9475121,10075480,9210812}) focus on FL within wireless networks, others (e.g., \cite{9141214,8917592,9120649,10857305}) explore FL as a tool for optimizing network operations. In this paper, we adopt the latter perspective, leveraging FL for distributed decision-making and resource allocation.\vspace*{-.3cm}

\subsection{Contribution}
Building upon the aforementioned references, this paper proposes a robust federated deep learning framework for joint communication and computation resource management under channel and computing uncertainties.
To this end, the paper considers a multi-BS empowered XR network, where the XR devices rely on strong edge computing capabilities (at the BSs) and high-quality communication links. Unlike our conference version \cite{10639072}, which considers only a single BS, this paper extends the framework to a multi-BS cellular XR network. Our objective is to minimize the worst-case delay across all devices by optimizing transmit power, computing power, and computation allocations. However, this problem is inherently challenging due to the stochastic constraints from channel and computing uncertainties, the communication and computation resource interdependency, and the network's distributed nature. Compared to \cite{10639072}, our extended model introduces additional operational challenges, such as multi-BS interference, scalable network-wide optimization, and the unique constraints of XR applications, all of which are considered in the current paper.

More specifically, to enable real-time decision-making, we employ an ML-based approach. Our proposed DNN takes estimated channels and computing requirements as inputs and outputs optimized power and computation allocations. The DNN can then be trained in an unsupervised manner using either centralized learning or an FL approach, where local BSs compute gradient updates that are aggregated at a central server to form a global model. A key novelty is the use of \emph{joint uncertainty injection}, where sampled channel and computing errors are injected directly into the DNN's output. 
\begin{figure*}[!t]
	\centering
	\includegraphics[width=.8\linewidth]{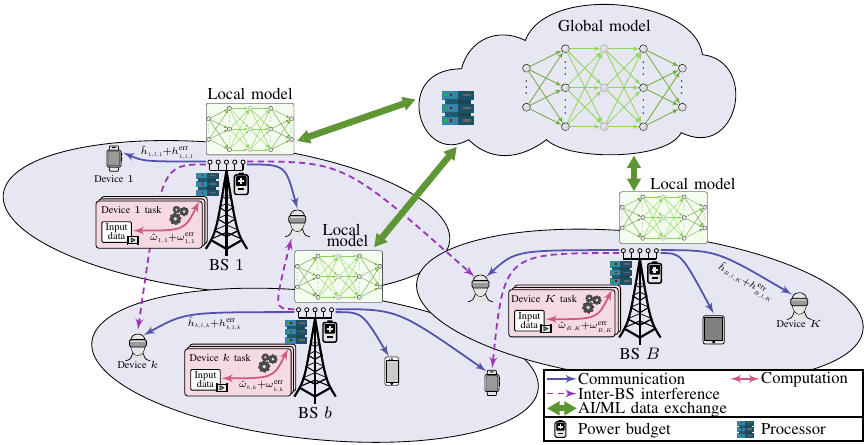}\\
	\vspace*{-.2cm}
	\caption{Network setup with one central server and global DNN model, $3$ multi antenna BSs each serving a number of XR devices accommodating a local DNN model. Devices' communication, computation, and inter-BS interference links are shown.}
	\label{fig:sys_mdl}
	\vspace*{-.5cm}
\end{figure*}

Extensive numerical evaluations validate the effectiveness of uncertainty injection in enhancing the network-wide robust delay performance across various uncertainty levels and network densities while maintaining scalability. The proposed FL-based solutions perform close to the centralized upper bound while significantly outperforming non-ML approaches, and offer practical benefits in terms of complexity and data exchange. The results demonstrate that joint uncertainty injection generalizes beyond single-uncertainty state-of-the-art schemes. {These findings underscore the feasibility of the proposed AI-driven, multi-purpose resource management framework for 
	real-world cellular XR applications.}
The key contributions are summarized as follows:
\begin{itemize}
	\item 
	We consider a cellular XR-based network with multiple distributed, edge-enabled BSs serving delay-sensitive XR devices for both computation (e.g., scene modeling, rendering) and communication (data transmission), under channel and computing uncertainties.
	\item 
	We formulate uncertainty-aware worst-case delay minimization problems under both centralized and distributed settings so as to determine the optimal transmit power, computing power, and computation allocations.
	\item 
	To overcome the complexity of conventional optimization, we develop robust centralized and federated deep learning frameworks, incorporating joint uncertainty injection (injecting channel and computing estimation errors after the DNN outputs) during training.
	\item 
	Our numerical results show that joint uncertainty injection improves the network-wide worst-case delay among XR devices. The proposed FL scheme approaches the centralized upper bound, outperforming non-ML baselines, while being computationally efficient, reducing data exchange, and refining the DNN in early training epochs. This highlights the proposed scheme's potential in provisioning joint communication and computing for robust XR networks.
\end{itemize}\vspace*{-.25cm}

\subsection{Notation and Organization}
We use boldface lowercase for vectors (e.g., $\mathbf{h}$), boldface uppercase for matrices (e.g., $\mathbf{H}$), and calligraphic letters for sets (e.g., $\mathcal{H}$). $\mathcal{F}_{\boldsymbol{\Theta}}$ denotes a DNN with parameter set $\boldsymbol{\Theta}$. Further, Hermitian transpose, transpose, absolute value, and $\ell_p$-norm are denoted as $\left( \cdot \right)^{\dagger}$, $\left( \cdot \right)^{T}$, $\left|\cdot\right|$, and $\norm[]{\cdot}_{p}$, respectively.

The paper is organized as follows: Section~\ref{sec:sysmod} presents the system components, XR operations, and joint communication and computing model, along with the uncertainty-aware worst-case delay minimization problem. Section~\ref{sec:dnn} introduces the centralized DNN and the joint uncertainty injection mechanism. Section~\ref{sec:fl} extends the discussion to the decentralized FL framework. Simulation results and analysis are shown in Section~\ref{sec:sim}, and the paper is concluded in Section~\ref{sec:con}.

\section{System Model and Problem Formulation}\label{sec:sysmod}\vspace*{-.05cm}
In this work, we consider an XR network comprising $B$ BSs equipped with embedded computing capabilities, collectively serving a total of $K$ single-antenna XR devices. Each BS is equipped with $L_b$ antennas and hosts a local ML model server to facilitate local training, parameter storage, and inference. We denote the set of BSs by $\mathcal{B}=\{1,\cdots,B\}$ and the set of XR devices by $\mathcal{K}=\{1,\cdots,K\}$. Furthermore, for BS $b$, the set of antennas is denoted by $\mathcal{L}_b=\{i_{1,b},\cdots,i_{L_b,b}\}$, with $i_{j,b}$ being the $j$-th antenna of the $b$-th BS, with $L$ being the combined number of all BSs' antennas. The network further comprises of a central processor (CP) with a global ML model server and processing capabilities. Depending on the CP's integration level and functionality, the network architecture may be categorized as multi-cell, coordinated (Section~\ref{sec:dnn}), or distributed (Section~\ref{sec:fl}) radio access network (RAN) \cite{10578301}. 

We assume disjoint XR device-clusters in this work, where each XR device $k$ is associated with exactly one BS. The set of devices served by BS $b$ is denoted by $\mathcal{K}_b\subseteq\mathcal{K}$, with $\mathcal{{K}}_b \cap \mathcal{{K}}_{b'}  = \emptyset, \forall b\neq b'$. Here, $K_b$ denotes the number of devices served by BS $b$. Fig.~\ref{fig:sys_mdl} illustrates the system model with $3$ BSs and $8$ XR devices, depicting communication, computation, and inter-BS interference links. In the centralized approach, the CP coordinates all BSs and hosts the network-wide DNN. In the FL-based approach, however, each BS trains a local model for its associated users and the CP maintains the global model, while each user is still primarily served by a single BS.\vspace*{-.2cm}

\subsection{XR Device Operation}
In many key XR applications, such as image and video processing or digital twinning, devices increasingly rely on robust remote processing capabilities provided by the network edge \cite{9363888}. For example, \cite{9615112} introduces \emph{UAVideo}, a framework in which aerial vehicles generate video data, process it locally or at the edge, and then forward it to devices. Minimizing the total delay in such frameworks is critical to ensuring smooth and responsive user experiences.
Each device's total delay comprises two primary components, namely the communication delay, representing the time incurred on the physical communication medium, and the computation delay, denoting the time required to complete the device's computational tasks.

{In our setup, devices such as head-mounted displays, tablets, smartphones, and smartwatches require advanced computation services provided by the network infrastructure. For instance, a head-mounted display may receive a 360-degree video stream pre-rendered and encoded at the BS. Concurrently, control inputs from the XR device, such as movement data or field-of-view changes, influence the rendering model. This creates a strong interdependence between modeling, pre-rendering, and encoding processes at the BS \cite{9363888}. Importantly, the computational term in our model captures these rendering and encoding processes, which are known to dominate the processing complexity in XR systems \cite{mangiante2017vr}. In contrast, control signaling involves significantly smaller data rates and latency contributions compared to the video stream \cite{10033423,10328648,mangiante2017vr}, and is, therefore, negligible relative to the dominant computation and communication delays. Similarly, cache access delays are typically much smaller than rendering and transmission delays in edge-assisted XR systems and are thus not explicitly modeled \cite{10033423}. Overall, this abstraction focuses on the dominant latency components, highlighting the interplay between communication and computing, and emphasizing the need for efficient resource allocation to meet the stringent latency and reliability requirements of XR applications.}\vspace*{-.2cm}

\subsection{Channel and Communication Delay Models}
Let $h_{b,l,k}\in\mathbb{C}$ be the channel coefficient from the $b$-th BS's antenna $l$ to device $k$, where $\bm{h}_{b,k} = [h_{b,i_{1,b},k},\cdots,h_{b,i_{L_b,b},k}]^T$ is the aggregate channel vector between BS $b$ and device $k$. Device $k$'s full aggregate channel vector including all BS links is denoted by $\bm{h}_{k} = [\bm{h}_{1,k}^T,\cdots,\bm{h}_{B,k}^T]^T$. We assume that the BSs only have access to an estimate of the channels. That is, the estimated channel from BS $b$'s antenna $l$ to device $k$ is denoted by $\hat{h}_{b,l,k}\in\mathbb{C}$, with the aggregate estimated channel vectors denoted by $\bm{\hat{h}}_{b,k} = [\hat{h}_{b,i_{1,b},k},\cdots,\hat{h}_{b,i_{L_b,b},k}]^T$ and $\bm{\hat{h}}_{k} = [\bm{\hat{h}}_{1,k}^T,\cdots,\bm{\hat{h}}_{B,k}^T]^T$. The estimated channels approximate the real channels only up to some extend, with some uncertainty consideration. The related literature provides a multitude of different uncertainty models, e.g., including additive or multiplicative uncertainties \cite{6169188}, assuming specific uncertainty bounds \cite{6036193}, or opting for statistical uncertainties \cite{10071987}.
In practice, the channel uncertainty distribution is not known \cite{6036193}. Yet, as channel estimations can be sampled in large amounts, a data-driven approach is able to \emph{learn} the channel uncertainty implicitly. Similar to the existing literature \cite{10071987}, we model the channel coefficients in our paper as follows:\vspace*{-.1cm}
\begin{equation}
	h_{b,l,k} = \hat{h}_{b,l,k} + h^\text{err}_{b,l,k},\label{eq:hlk}\vspace*{-.1cm}
\end{equation}
where $h^\text{err}_{b,l,k}\sim\mathcal{CN}(0,\sigma_h^2)$ is the channel estimation error. {Note that the uncertainty-aware framework adopted in \eqref{eq:hlk} does not require explicit knowledge of the underlying distribution of $h^\text{err}_{b,l,k}$. In practice, this distribution may vary depending on the channel estimation process (e.g., Gaussian, Rician, or mixed fading statistics). The proposed algorithm operates purely on sampled realizations of the uncertainty and is, therefore, agnostic to the specific parametric form of the distribution. For the sake of reproducibility and consistency with prior work, the simulations in this paper adopt the Gaussian assumption, which is widely used in the literature, e.g., \cite{10071987} and references therein. In Sec.~\ref{ssec:distribution}, we further extend the evaluation to non-Gaussian uncertainty models to validate the generalization capability of the proposed framework.}

The beamforming vector for device $k$, transmitted by BS $b$, is denoted by $\bm{v}_{b,k}\in\mathbb{C}^{L_b}$ and defined as $\bm{v}_{b,k} = [v_{b,i_{1,b},k},\cdots,v_{b,i_{L_b,b},k}]^T$, where 
$\norm[]{\bm{v}_{b,k}}_2 = 1$. The aggregate beamforming vector for device $k$ is represented as $\bm{v}_{k} = [\bm{v}_{1,k}^T,\cdots,\bm{v}_{B,k}^T]^T$.
Since each device is associated with only one BS, $\bm{v}_{k}$ is group-sparse by design. This sparsity is enforced by setting $\bm{v}_{b,k} = \bm{0}^{L_b}$ for all $k\in \mathcal{K}\backslash\mathcal{K}_b$, at each BS $b$. 
Furthermore, let $p_{k}^\text{tx}$ denote the transmit power allocated by the serving BS to device $k$. The transmit power vector is then represented as $\bm{p}^\text{tx} = [p_{1}^\text{tx},\cdots,p_{K}^\text{tx}]^T$.

{It is important to note that, due to the different time scales between transmit power or computational resource allocation and the device-to-transmitter association, a fixed association provides a reasonable modeling abstraction within the scope of this paper. In practice, association and handover decisions occur at slower time scales and can be treated as outer-loop processes \cite{1603708}. Moreover, any valid association satisfying the aforementioned constraints can be incorporated without fundamentally changing the proposed framework. For moderate variations in the active device set, this can be achieved using standard techniques such as zero-padding, masking, or maximum-size system design \cite{8633948}. Consequently, the presented results are indicative of performance under different underlying associations.}

{To clarify the time scales discrepancy among the considered variables, we consider three different categories, namely, \emph{radio scheduling/inference} (fast), \emph{channel/workload variations} (medium), and \emph{model training} (slow). Communication time slots are typically in the order of sub-ms \cite{etsiNR} and XR rendering budgets are around $10\,$ms per frame, with DNN inference running once per BS per scheduling interval. In contrast, channel statistics and XR workload profiles evolve in the higher ms-range (mobility) up to seconds (XR scene changes). Lastly, DNN training and FL aggregation occur offline or in the background at a slower time scale (minutes-hours) and are not part of the real-time loop; only inference is latency-critical \cite{10239349}. This decoupling ensures that only the lightweight inference operates on the sub-ms/ms budget, while (re)training can be triggered when long-term statistics, such as user density or mobility patterns, change \cite{8227766}. Extensions toward fully dynamic association and variable-size architectures (e.g., permutation-invariant or graph-based models) are left for future work.}


The transmitted messages for all devices are encoded as $\xi_k$, $\forall k\in\mathcal{K}$, where $\xi_k\sim\mathcal{CN}(0,1)$ are independent and identically distributed, circularly symmetric Gaussian random variables. Each device $k$ receives its own desired signal, interference from other signals, and noise. Device $k$'s received signal is\vspace*{-.2cm}
\begin{align}
y_k = \bm{h}_{k}^H \bm{v}_{k} \sqrt{p^\text{tx}_k} \xi_k + \sideset{}{_{j\in\mathcal{K}\backslash\{k\}}}{\sum} \bm{h}_{k}^H \bm{v}_{j} \sqrt{p^\text{tx}_j} \xi_j + n_k, \label{eq:yk}\\[-.75cm]\nonumber
\end{align}
where $n_k$ represents the additive white Gaussian noise with zero mean and variance $\sigma^2$. 
The achievable rate of device $k$ is then given by\vspace*{-.2cm}
\begin{align}
r_k = W\log_2\left(1+\frac{|\bm{h}_k^H \bm{v}_k|^2 p^\text{tx}_k}{\varphi_k + \sigma^2}\right),\label{eq:rk}\\[-.65cm]\nonumber
\end{align}
with system bandwidth $W$ and total interference 
\begin{align}
\varphi_k &= \underbrace{\sum_{j\in\mathcal{K}_b\backslash\{k\}} |\bm{h}_k^H \bm{v}_j|^2 p^\text{tx}_j}_{\varphi_k^\text{intra}} + \underbrace{\sum_{j\in\mathcal{K}\backslash\{\mathcal{K}_b\}} |\bm{h}_k^H \bm{v}_j|^2 p^\text{tx}_j}_{\varphi_k^\text{inter}},\label{eq:varphik}\\[-.7cm]\nonumber
\end{align}
where $b$ denotes the BS serving device $k$, while $\varphi_k^\text{intra}$ and $\varphi_k^\text{inter}$ represent the intra-BS and inter-BS interference levels, respectively. These terms are detailed as follows:
\begin{itemize}
\item \emph{Intra-BS interference}: This interference arises from devices served by the same BS as device $k$. This interference can be managed by the BS through power allocation.
\item \emph{Inter-BS interference}: This interference originates from devices served by other BSs and is beyond the control of device $k$'s serving BS.
\end{itemize}\vspace*{-.3cm}

\subsection{Computing Framework and Computation Delay}
From the computation perspective, each XR device requires the execution of specific computational tasks, such as field-of-view pre-rendering, object tracking, or activity detection \cite{10033423}. In this work, device $k$ requests to process a task, which consists of an input data size $D_k^\text{in}$, and a computation intensity $\omega_k$, i.e., the required computation cycles per bit of input data is denoted by $\omega_k$, for a total of $\omega_kD^{in}_k$ computation cycles for device $k$'s task. However, such a computational requirement $\omega_k$ may not be perfectly known at the corresponding BS, as only an estimation $\hat{\omega}_k$ is available.
Specifically, computing environments are highly dynamic, i.e., they vary depending on system load and concurrent processes, and systems dependency, i.e., the inter-relation of input data characteristics \cite{yuan2003energy}.
Various statistical models for $\omega_k$ exist in the literature, including uniform, normal, and Gamma distributions \cite{lorch2001improving}. 
This paper handles computation uncertainty in a data-driven manner by training on a large set of sampled uncertainty realizations. The computation intensity is modeled as\vspace*{-.1cm}
\begin{equation}
\omega_{k} = \hat{\omega}_{k} + \omega^\text{err}_{k},\label{eq:omegak}
\end{equation}
where $\omega^\text{err}_{k}\sim\mathcal{N}(0,\sigma_\omega^2)$ is the computing estimation error. Similar to the Gaussian assumption remark made earlier about the channel estimation error, we also note here that the computation estimation error is assumed to be Gaussian for simulations purposes. The framework of our paper is, however, agnostic to the underlying error distribution, since the algorithm operates solely on sampled error realizations.

With an estimation of the required computation cycles for device $k$'s task, i.e., $\hat{\omega}_k D_k^\text{in}$, the corresponding BS allocates $f^\text{co}_k$ in cycles/s to compute $k$'s task. The overall vector of computation allocations is denoted by $\bm{f}^\text{co} = [f^\text{co}_{1},\cdots,f^\text{co}_{K}]^T$. Due to the finite computation resources at the BSs, the maximum computation capacity constraint is given by\vspace*{-.1cm}
\begin{equation}
\sideset{}{_{k\in\mathcal{K}_b}}\sum f^\text{co}_{k} \leq F_b^{\text{max}}, \qquad \forall b\in\mathcal{B}, \label{eq:maxcompcap}
\end{equation}
where $F_b^\text{max}$ is BS $b$'s maximum computation capacity. Due to the intertwined nature of communication and computation resources, especially in terms of the allocated power levels, we consider $p_b^\text{co}$ as the total power allocated by BS $b$ for the computation of all tasks. The overall vector of all BSs computation powers is denoted by $\bm{p}^\text{co} = [p^\text{co}_{1},\cdots,p^\text{co}_{B}]^T$. The relation between the computation power $p_b^\text{co}$ and computation allocation $f_k$ is given by\vspace*{-.2cm}
\begin{equation}
p_b^\text{co} = \tau \left(\sideset{}{_{k\in\mathcal{K}_b}}\sum f^\text{co}_{k} \right)^{\mu} \label{eq:pco},
\end{equation}
where $\tau$ and $\mu$ are constants of the CPU model \cite{10328648}.\vspace*{-.3cm}

\subsection{Joint Communication and Computation Considerations}\vspace*{-.1cm}
By jointly considering the communication and computation delay, device $k$ experiences a total delay $t_k$ as follows\vspace*{-.15cm}
\begin{align}
t_k = \frac{\omega_k D_k^\text{in}}{f^\text{co}_k} + \frac{D_k^\text{out}}{r_k}, \label{eq:tk}\\[-.7cm]\nonumber
\end{align}
where $D_k^\text{out}$ denotes the output data size of device $k$'s task computation\footnote{In practice, channel fluctuations occur on sub-ms scales, while workload variations evolve over tens of ms to seconds. Both are here abstracted as uncertain system states in \eqref{eq:tk}, since they jointly affect the end-to-end XR delay. This modeling choice enables a unified robust optimization framework.}. While $\omega_k$ in \eqref{eq:tk} directly reflects computation uncertainty (see \eqref{eq:omegak}), the rate $r_k$ depends on the channel conditions (see \eqref{eq:rk}), thus capturing the impact of channel uncertainty. 
In order to achieve fair service among all devices, the worst-case delay is computed as $t_\text{max} = \text{max}_{k\in\mathcal{K}}\{t_k\}$.

In this work, we consider the coupling of communication and computation power consumption (i.e., $\bm{p}^\text{tx}$ and $p_b^\text{co}$, respectively) at the BSs, which leads to the overall power constraint\vspace*{-.15cm}
\begin{align}
\sideset{}{_{k\in\mathcal{K}_b}}\sum p_k^\text{tx} + p_b^\text{co} \leq P_b^\text{max}, \qquad \forall b\in\mathcal{B},\label{eq:P}\\[-.7cm]\nonumber
\end{align}
where $P_b^\text{max}$ is the maximum allowable power per BS \cite{8330764}. Eq. \eqref{eq:P} emphasizes the intertwined communication and computation powers, especially underlining the need for a careful selection of joint resource allocation under limited resources.\vspace*{-.2cm}


\subsection{Robust Delay Utility}\vspace*{-.1cm}
Based on \eqref{eq:tk}, one can infer that the overall delay $t_k$ of device $k$ is a function of the channel vectors $\bm{h}_k$ and the computation intensities $\omega_k$, where both $\bm{h}_k$ and $\omega_k$ follow some unknown probability distributions. Specifically, each BS only has access to the estimated $\bm{\hat{h}}_k$ and $\hat{\omega}_k$, $\forall k\in\mathcal{K}_b$.

To best account for the channel and computing uncertainties, the paper utilizes a DNN with inputs $\bm{\hat{h}}_k$, $\hat{\omega}_k$, $\forall k\in\mathcal{K}$, and inject samples of uncertainty realizations of both channels and computing requirements into the training phase just after the DNN output \cite{10071987}. That is, for a given resource allocation, we inject estimation errors to impact the objective value, and afterwards train the network according to the updated objective.
Such uncertainty injection supplies a multitude of potential worst-case delays $t_\text{max}$ for different realizations. This paper then aims at minimizing the $\gamma$-th quantile of the worst-case delay distribution, denoted by $t_\text{max}^\gamma$, where\vspace*{-.15cm}
\begin{align}
\text{Pr}\left[t_\text{max} > t_\text{max}^\gamma \big| \bm{\hat{h}}_k, \hat{\omega}_k, \forall k\in\mathcal{K} \right] \leq \gamma,\label{eq:t_percentile}\\[-.75cm]\nonumber
\end{align}
where $t_\text{max}^\gamma$ can be interpreted as the value for which only a fraction $\gamma$ of all realizations have higher worst-case delays.

While we adopt the robust delay utility in \eqref{eq:tk} to capture the joint impact of channel and computation uncertainties, other robustness approaches can be found in the existing literature. For instance, common ML-based techniques include data augmentation \cite{8936957} and noise injection \cite{shaham2018understanding}, where uncertainty is artificially added to the training data or feature space. While effective in some domains, these methods do not directly capture the system-level uncertainties of communication and computation considered in the current paper. From an optimization perspective, robust designs based on sample average approximation (SAA) combined with weighted minimum mean square error (WMMSE) algorithms have been proposed (e.g., \cite{reifert2021rate}) to address channel uncertainty. Such methods, however, are computationally demanding and less suitable for real-time operation. In contrast, our uncertainty injection scheme integrates robustness directly into the DNN training process, enabling the model to implicitly learn the uncertainty distribution \cite{10071987}.\looseness-1
\vspace*{-.25cm}

\subsection{Problem Formulation}\vspace*{-.1cm}
The paper aims at minimizing the robust worst-case delay $t_\text{max}^\gamma$ among all devices. The corresponding optimization problem can then be mathematically formulated as follows:\vspace*{-.2cm}
\begin{subequations}\label{eq:Opt1}
\begingroup
\addtolength{\jot}{-.1cm}
\begin{align}
	\underset{\bm{p}^\text{tx},\bm{p}^\text{co},\bm{f}^\text{co}}{\text{min}}\quad &t_\text{max}^\gamma  \tag{\ref{eq:Opt1}} \\
	\text{s.t.} \quad\quad & \eqref{eq:hlk},\eqref{eq:omegak},\eqref{eq:maxcompcap},\eqref{eq:P}, \eqref{eq:t_percentile},\hspace*{-.2cm} \nonumber\\
	&t_\text{max} \geq t_k, &\forall k\in\mathcal{K},\label{eq:tmaxconstraint}\\[-.65cm]\nonumber
\end{align}
\endgroup
\end{subequations}
where 
the computation capacity constraint is given in \eqref{eq:maxcompcap}, and the power constraint is given in \eqref{eq:P}. Constraint \eqref{eq:t_percentile} ensures the solution's robustness in terms of the $\gamma$-th quantile. Constraint \eqref{eq:tmaxconstraint} expresses the worst delay among all devices as $t^\text{max}$.
The intricacy of problem \eqref{eq:Opt1} stems from the mathematical coupling between the communication and computation variables and constraints, as well as from the unknown distribution of the worst-case delay $t_\text{max}$ given the channel and computing states uncertainties. 
Additionally, another challenge of problem \eqref{eq:Opt1} lies in its centralized nature. Specifically, solving problem \eqref{eq:Opt1} requires global knowledge of all channel vectors and computation requirements, $\bm{\hat{h}}_k$ and $\hat{\omega}_k$, $\forall k\in\mathcal{K}_b$, $\forall b\in\mathcal{B}$

The paper next proposes a data-driven solution approach based on a deep learning framework with joint uncertainty injection in Section~\ref{sec:dnn}. Afterwards, Section~\ref{sec:fl} presents the distributed federated learning-based implementation of the proposed joint uncertainty-based deep learning protocol. 

\section{Uncertainty-Aware Centralized Learning}\label{sec:dnn}
To tackle the intricate optimization problem \eqref{eq:Opt1}, this section harnesses the promising abilities of deep learning in terms of automated learning, feature representation, and versatility. Specifically, we utilize a DNN to directly map the problem parameters to a resource allocation solution, which is then further processed by injecting uncertainties for optimizing a robust objective.
In a typical DNN, neurons are organized into layers, including input, hidden, and output layers. These layers are densely interconnected, with each connection between neurons having an associated weight that is continuously adjusted during the training process. Additionally, nonlinear activation functions are applied to the neurons within these layers. In general, depth (number of layers) and width (number of neurons in each layer) of the DNN are key factors that influence its capacity to learn and represent data \cite{8755300}.

In a centralized manner, all BSs are connected to and coordinated by a CP with a global ML model server and processing capabilities. We propose a DNN architecture, which takes the estimated channels and estimated computing requirements as inputs, i.e., $\bm{\hat{h}}_k$ and $\hat{\omega}_k$, $\forall k\in\mathcal{K}$. The input layer consists of $2LK+K$ neurons, which represent the $2LK$ real and imaginary channel coefficients and the $K$ distinct devices' tasks. After the input layer, we attach $\eta_h$ fully-connected hidden layers with $n_h$ neurons each, and Rectified Linear Unit (ReLU) activation functions. The output layer consists of $2K+B$ neurons, where the first $K+B$ entries represent the $K$ devices' allocated power and the BSs' computation powers, and the remaining $K$ entries represent the computation capacities. The DNN then returns the output vector $\bm{x}=[x_1,\cdots,x_{2K+B}]^T$, representing the optimized resource allocation $\bm{p}^\text{tx}$, $\bm{p}^\text{co}$, and $\bm{f}^\text{co}$, after some mathematical manipulations. 
We let $\mathcal{F}_{\boldsymbol{\Theta}}$ denote the proposed DNN, such that problem \eqref{eq:Opt1} is solved by\vspace*{-.15cm}
\begin{align}
\bm{x} = \mathcal{F}_{\boldsymbol{\Theta}}(\bm{\hat{h}},\bm{\hat{\omega}}), \label{eq:dnn}\\[-.75cm]\nonumber
\end{align}
where $\bm{\hat{h}}=[\bm{\hat{h}}_{1}^T,\cdots,\bm{\hat{h}}_{K}^T]^T$ and $\boldsymbol{\hat{\omega}}=[\hat{\omega}_{1},\cdots,\hat{\omega}_{K}]^T$. Note that $\boldsymbol{\Theta}$ refers to the DNN parameters, e.g., weights and biases.\footnote{In equation \eqref{eq:dnn}, the output $\bm{x}$ is a function of the DNN model parameters $\boldsymbol{\Theta}$. Such DNN $\mathcal{F}_{\boldsymbol{\Theta}}$ and its parameters $\boldsymbol{\Theta}$ are sequentially updated via backpropagation as discussed in Section \ref{ssec:backprop} later.}
\begin{figure}[!t]
\centering
\includegraphics[width=\linewidth]{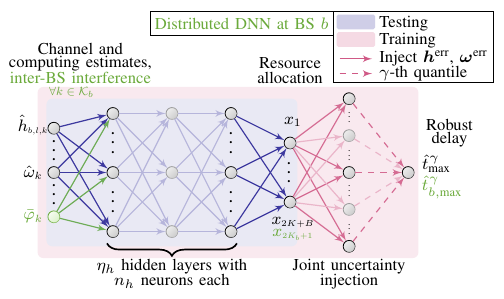}\\
\vspace*{-.3cm}
\caption{Proposed DNN architecture with estimated input, resource allocation output, and uncertainty injection. Modifications for the distributed DNN are highlighted in green.}
\label{fig:dnn_mdl}
\vspace*{-.5cm}
\end{figure}
Fig.~\ref{fig:dnn_mdl} illustrates a representation of the proposed DNN-based solution. 

In the context of minimizing the robust worst-case delay among devices, we next explore how to map the DNN output to the resource allocation, how to ensure the robustness of the solution, i.e., in terms of the $\gamma$-th quantile guarantee, and how to train the DNN according to the robust objective.\vspace*{-.25cm}

\subsection{DNN-Driven Resource Allocation Strategy}\vspace*{-.05cm}
A major complication of the joint communication and computation resource management problem \eqref{eq:Opt1} is the interdependent computation capacity \eqref{eq:maxcompcap} and power \eqref{eq:P} constraints.
We, thus, first use the softmax function\footnote{The softmax function of an input vector $\bm{x}$ is an output vector, the $i$-th entry of which is $e^{x_i}/(\sum_{j=1}^{n}e^{x_j})$, where $n$ is the dimension of $\bm{x}$.} on parts of the DNN output to compute each BS's transmit and computation powers for all $b\in\mathcal{B}$ by computing \vspace*{-.05cm}
\begin{equation}
[(\bm{p}_b^\text{tx})^T,p_b^\text{co}]^T = P_b^\text{max}\cdot \text{softmax}\big([x_{j^1_b},\cdots,x_{j^{K_b+1}_b}]^T\big),\label{eq:ptx_pco}
\end{equation}
where the indices $j^1_b$ and $j^{K_b+1}_b$ denote the first and last output associated with BS $b$, respectively. We can calculate them as $j^1_b = (b-1)(K_b+1)+1$ and $j^{K_b+1}_b = b(K_b+1)$, e.g., for BS $1$, $j^1_1 = 1$ and $j^{K_b+1}_1 = K_b+1$, respectively\footnote{Note that the calculation of $j^1_b$ and $j^{K_b+1}_b$ is based on the assumption that all BS have the same number of devices, i.e., $K_b = K_{b'}$, $b\neq b'$. While the proposed framework does not necessarily require all BS to have the same number of devices, such case requires a different management of indices, e.g., using a lookup table.}.
By using the softmax function, we 
ensure the full utilization of the complete power budget at each BS. Next, we transform the computation power into cycles using \eqref{eq:pco}, so as to obtain the maximum number of available computation cycles limited by the power allocation, denoted by $F_b^\text{pow}$. Based on \eqref{eq:pco}, $F_b^\text{pow}$ can be written as:\vspace*{-.1cm}
\begin{equation}
F_b^\text{pow} = \sqrt[\leftroot{1}\uproot{1}\mu]{p_b^\text{co}\,\tau^{-1}}.
\end{equation}
At this point, there are two limiting factors for the computation allocation (i.e., $\bm{f}^\text{co}$), namely, the BSs' maximum computation capacities $F_b^\text{max}$ in \eqref{eq:maxcompcap}, and the above power-limited available computation cycles $F_b^\text{pow}$, $\forall b\in\mathcal{B}$. We then use the smallest among the two limiting factors, i.e., $\text{min}\big(F_b^\text{max},F_b^\text{pow}\big)$. To determine the computation allocation $\bm{f}_b^\text{co}$, we use the softmax function on remaining parts of the output layer $\forall b\in\mathcal{B}$ by
\begin{equation}
\bm{f}_b^\text{co} = \text{min}\big(F_b^\text{max},F_b^\text{pow}\big)\cdot \text{softmax}\big([x_{i_b^1},\cdots,x_{i_b^{K_b}}]^T\big),\label{eq:fco}
\end{equation}
which utilizes either the capacity- or power-limited computation cycles. Here, the indices $i^1_b$ and $i^{K_b}_b$ can be calculated as $i^1_b = K+B+(b-1)K_b+1$ and $i^{K_b}_b = K+B+bK_b$.\vspace*{-.3cm}

\subsection{Joint Uncertainty Injection Scheme}\label{ssec:IIIb}\vspace*{-.05cm}
Given the resource allocation $\bm{p}^\text{tx}$, $\bm{p}^\text{co}$, and $\bm{f}^\text{co}$ found via \eqref{eq:ptx_pco} and \eqref{eq:fco}, one can readily compute a worst-case delay for \emph{one given} channel and computing estimation. However, problem \eqref{eq:Opt1} aims at providing a robust $t_\text{max}^\gamma$ accounting for the uncertainties $\bm{h}^\text{err}=[{h}_{1,1,1}^\text{err},\cdots,{h}_{B,L_B,K}^\text{err}]^T$ and $\boldsymbol{\omega}^\text{err}=[\omega_1^\text{err},\cdots,\omega_K^\text{err}]^T$, given in \eqref{eq:hlk} and \eqref{eq:omegak}, respectively. 
Thus, we next explain the uncertainty injection approach as first proposed by \cite{10071987}, and set forth the joint uncertainty injection scheme adopted in the joint computation and communication context considered in the current paper.

The basic idea of uncertainty injection is to estimate the statistical objective numerically, using a large number of realizations or samples of the uncertainty. Thereafter, such realizations are \emph{injected} after the DNN output to compute a large number of objective values. These objectives are then used to obtain an empirical estimate of the robust objective. Such estimate is then used to derive gradients and update the DNN parameters via backpropagation \cite{goodfellow2016deep}.

In this paper, however, the problem is subject to both channel and computing state information uncertainties, as reflected in equations \eqref{eq:hlk} and \eqref{eq:omegak}. The adopted objective $t_\text{max}^\gamma$ in problem \eqref{eq:Opt1} is, however, a function of the $\gamma$-th quantile of the worst-case delay distribution, as defined in \eqref{eq:t_percentile}. The network only has access to the estimated channel states, $\bm{\hat{h}}$, and computing requirements, $\bm{\hat{\omega}}$, while the true worst-case delay distribution depends on the unknown channel and computing error distributions, $\bm{h}^\text{err}$ and $\bm{\omega}^\text{err}$, respectively. To address this uncertainty, we adopt a sampling-based approach. 

As shown in Fig.~\ref{fig:dnn_mdl}, the DNN operates on estimated channel and computing requirement inputs, $\bm{\hat{h}}$ and $\bm{\hat{\omega}}$. After propagating through multiple hidden layers, the output vector is obtained as $\bm{x}=\mathcal{F}_{\boldsymbol{\Theta}}(\bm{\hat{h}},\bm{\hat{\omega}})$. Using equations \eqref{eq:ptx_pco} and \eqref{eq:fco}, the corresponding resource allocations are derived. At this stage, the BSs \emph{sample} $N$ realizations of the channel and computing uncertainties, denoted by $\bm{h}^\text{err}_n$ and $\bm{\omega}^\text{err}_n$, for $n\in{1,\cdots,N}$. These samples allow the network to compute $N$ worst-case delays, given by $t_{\text{max},n}$, $\forall n\in{1,\cdots,N}$ as follows\vspace*{-.15cm}
\begin{align}\label{eq:tmaxn}
t_{\text{max},n} = \text{max}_{k\in\mathcal{K}}\bigg\{\frac{(\hat{\omega}_k+\omega^\text{err}_{k,n}) D_k^\text{in}}{f^\text{co}_k} + \frac{D_k^\text{out}}{r_{k,n}}\bigg\},\\[-.7cm]\nonumber
\end{align}
where $r_{k,n}$ represents the data rate of the $k$-th XR device as a function of channel uncertainty, computed by incorporating \eqref{eq:hlk} into \eqref{eq:rk} as\vspace*{-.15cm}
\begin{align}
r_{k,n} = W\log_2\left(1+\frac{|\big(\bm{\hat{h}}_k+\bm{h}_{k,n}^\text{err}\big)^H \bm{v}_k|^2 p^\text{tx}_k}{\varphi_k + \sigma^2}\right).\\[-.6cm]\nonumber
\end{align}
This process, referred to as \emph{joint uncertainty injection} in Fig.~\ref{fig:dnn_mdl}, occurs immediately after the DNN output. The computed delay values are then sorted, and the $(\gamma\cdot N)$-th highest value is selected as an empirical estimate of the $\gamma$-th quantile of the worst-case delay, denoted as $\hat{t}_\text{max}^{\hspace*{.02cm}\gamma}$\footnote{A practical Python implementation of this step can be achieved using the \emph{quantile} function: $\hat{t}_\text{max}^{\hspace*{.02cm}\gamma} = \text{quantile}_{q=0.95}\big([t_{\text{max},1},\cdots,t_{\text{max},N}]\big)$.}.

It is worth noting herein that the above uncertainty injection procedure is applied \emph{only} during training, enabling the DNN to implicitly learn the underlying uncertainty distributions. Once trained, the DNN can robustly allocate resources based on estimated inputs without requiring further uncertainty sampling. As depicted in Fig.~\ref{fig:dnn_mdl}, training-related components are marked \emph{red}, while real-time operation and testing are highlighted \emph{blue}.\vspace*{-.4cm}

\subsection{Unsupervised Learning via Backpropagation}\label{ssec:backprop}\vspace*{-.05cm}
An integral part of the proposed solution is the DNN training procedure, especially in accounting for the uncertainties to ensure a robust solution to problem \eqref{eq:Opt1}. More specifically, by taking the $(\gamma\cdot N)$-th highest value (i.e., $\hat{t}_\text{max}^{\hspace*{.02cm}\gamma}$) of all computed worst-case delays $t_{\text{max},n}$, $\forall n\in\{1,\cdots,N\}$, gradients are derived from $\hat{t}_\text{max}^{\hspace*{.02cm}\gamma}$ to update the DNN parameters. To be more specific, we use the gradient $\partial\hat{t}_\text{max}^{\hspace*{.02cm}\gamma}/\partial\Theta$, which is computed using Pytorch \cite{paszke2019pytorch}. In that way, the DNN is trained in an unsupervised fashion, implicitly learning the distribution of channel and computing uncertainties. The numerical merit of our proposed method is further discussed in the simulations section of the paper, as shown in Section~\ref{sec:sim}. 

For completeness, the deep learning training procedure incorporating joint uncertainty injection is detailed in Algorithm~\ref{alg:CENtrain}. In summary, for each epoch, Algorithm~\ref{alg:CENtrain} partitions the training process into minibatches, iterating $M$ times with newly sampled realizations in each iteration. After processing all minibatches, the DNN's performance is validated, and model parameters are saved if they yield the lowest observed robust delay, i.e., $\hat{t}_{\text{max},\text{global}}^{\hspace*{.02cm}\gamma}$. Within each minibatch (and during validation), the CP obtains channel and computing requirement estimates, performs DNN inference, determines resource allocations for each network realization, and updates the DNN parameters based on the computed gradients. Note that the last step is applied only during training. This iterative process continues until either a predefined convergence criterion is met or when the maximum number of training epochs is reached.

\begin{algorithm}[h]
\caption{Proposed Centralized Training Procedure}
\begin{algorithmic}[1]
	\REPEAT
	\FOR{$j = 1,\cdots,M$}
	\STATE Obtain batch $\bm{\hat{h}}$ and $\bm{\hat{\omega}}$
	\STATE Compute DNN output vector $\bm{x}$ via \eqref{eq:dnn}
	\STATE Determine $\bm{p}^\text{tx}$, $\bm{p}^\text{co}$, and $\bm{f}^\text{co}$ via \eqref{eq:ptx_pco} and \eqref{eq:fco}
	\STATE Compute $\hat{t}_\text{max}^{\hspace*{.02cm}\gamma}$ as described in Section~\ref{ssec:IIIb}
	\STATE DNN model update using gradient $\partial\hat{t}_\text{max}^{\hspace*{.02cm}\gamma}/\partial\Theta$
	\ENDFOR
	\STATE Obtain \textit{validation} set $\bm{\hat{h}}$ and $\bm{\hat{\omega}}$
	\STATE Compute DNN output vector $\bm{x}$ via \eqref{eq:dnn}
	\STATE Determine $\bm{p}^\text{tx}$, $\bm{p}^\text{co}$, and $\bm{f}^\text{co}$ via \eqref{eq:ptx_pco} and \eqref{eq:fco}
	\STATE Compute $\hat{t}_\text{max}^{\hspace*{.02cm}\gamma}$ as described in Section~\ref{ssec:IIIb}
	\IF{$\hat{t}_\text{max}^{\hspace*{.02cm}\gamma} < \hat{t}_{\text{max},\text{global}}^{\hspace*{.02cm}\gamma}$}
	\STATE Save DNN parameters $\bm{\Theta}$ and set $\hat{t}_{\text{max},\text{global}}^{\hspace*{.02cm}\gamma} = \hat{t}_\text{max}^{\hspace*{.02cm}\gamma}$
	\ENDIF
	\UNTIL{Convergence}
\end{algorithmic}
\label{alg:CENtrain}
\end{algorithm}

Departing from the single-CP optimization problem \eqref{eq:Opt1} and its centralized solution via Algorithm~\ref{alg:CENtrain}, we next explore a practical distributed network architecture based on FL.\vspace*{-.2cm}

\section{Uncertainty-Aware Federated Learning}\label{sec:fl}\vspace*{-.05cm}
In this section, we consider the practically feasible and scalable distributed network architecture, as illustrated in Fig.~\ref{fig:sys_mdl}.
At the foundation of the decentralized network design, each BS independently performs model training, inference, and resource allocation. Under such realm, the CP serves solely as a global parameter server, aggregating and updating ML models without access to global network state information, such as channels, computation requirements, or interference levels. To this end, the next part of the paper formulates a decentralized optimization problem that is better suited to the practical constraints and scalability requirements of future large-scale systems, and proposes a FL-approach to solve the jointly robust optimization problem using a joint uncertainty injection mechanism.\vspace*{-.4cm}

\subsection{Distributed Network Architecture}\label{ssec:cendis}\vspace*{-.05cm}
In the centralized setup, the CP collects data from the BSs, trains a global ML model discussed in Section~\ref{sec:dnn}, and determines the resource allocation for each BS. The following proposed distributed network architecture, however, redefines the CP's role as a global parameter server responsible for ML model aggregation, updates, and transmission. Training and inference are, nevertheless, performed locally at each BS, as illustrated in Fig.~\ref{fig:sys_mdl}. This results in a hybrid RAN architecture that exhibits both distributed and coordinated characteristics \cite{10578301}. Specifically, resource allocation decisions are made locally at each BS, while the CP facilitates network-wide training and the exchange of interference information.

The distributed version of problem \eqref{eq:Opt1} for BS $b$ becomes\vspace*{-.2cm}%
\begin{subequations}\label{eq:Opt2}
\begingroup
\addtolength{\jot}{-.1cm}
\begin{align}
	\underset{\bm{p}_b^\text{tx},p_b^\text{co},\bm{f}_b^\text{co}}{\text{min}}\quad &t_{b,\text{max}}^\gamma  \tag{\ref{eq:Opt2}} \\
	\text{s.t.} \quad\quad & \eqref{eq:hlk},\eqref{eq:omegak},\eqref{eq:maxcompcap},\eqref{eq:P}, \hspace*{-.2cm} \nonumber\\
	&\text{Pr}\left[t_{b,\text{max}} > t_{b,\text{max}}^\gamma \big| \bm{\hat{h}}_k, \hat{\omega}_k, \forall k\in\mathcal{K}_b \right] \leq \gamma, \hspace*{-1.45cm}\label{eq:t_percentile_dist}\\
	&\varphi_k^\text{inter} = \sideset{}{_{j\in\mathcal{K}\backslash\{\mathcal{K}_b\}}}{\sum} |\bm{h}_k^H \bm{v}_j|^2 p_j^\text{tx},\hspace*{-.3cm} &\forall k\in\mathcal{K}_b,\label{eq:phi_inter}\\
	&t_{b,\text{max}} \geq t_k, &\forall k\in\mathcal{K}_b.\label{eq:tmaxconstraint_dist}\\[-.6cm]\nonumber
\end{align}
\endgroup
\end{subequations}
Here, the optimization variables $\bm{p}_b^\text{tx}$, $p_b^\text{co}$, and $\bm{f}_b^\text{co}$ correspond to the transmit power, computation power, and computation cycles allocated to devices associated with BS $b$. The transmit power vector $\bm{p}_b^\text{tx}$ and computation allocation vector $\bm{f}_b^\text{co}$ are subsets of $\bm{p}^\text{tx}$ and $\bm{f}^\text{co}$, respectively, specific to the devices in $\mathcal{K}_b$. The objective is to minimize the worst-case delay among the devices served by BS $b$, expressed as $t_{b,\text{max}} = \text{max}_{k\in\mathcal{K}_b}\{t_k\}$, with $t_{b,\text{max}}^\gamma$ denoting the $\gamma$-th quantile of the worst-case delay distribution. The constraint set of problem \eqref{eq:Opt2} is particularly augmented with constraint \eqref{eq:phi_inter}, which accounts for the inter-BS interference terms. Specifically, the decentralized network architecture requires careful management of inter-BS interference $\varphi_k^\text{inter}$, as discussed next.\vspace*{-.3cm}

\subsection{Channel Knowledge and Inter-BS Interference}
\begin{figure}[!t]
\centering
\includegraphics[width=\linewidth]{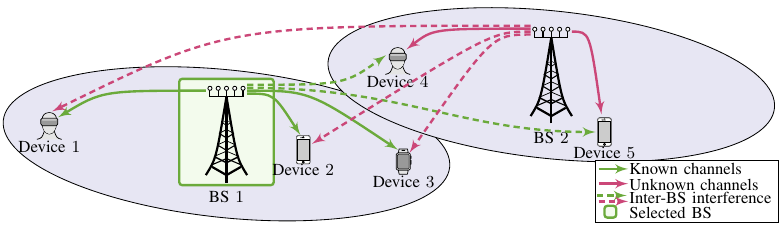}\\
\vspace*{-.2cm}
\caption{Channel knowledge representation for 2 BSs and 5 devices illustrating known and unknown channels.}
\label{fig:channel_knowledge}
\vspace*{-.6cm}
\end{figure}
Each device experiences both intra-BS and inter-BS interference, which influence the achievable rate in equation \eqref{eq:rk} and, consequently, the communication delay as a component of $t_k$ (see equation \eqref{eq:tk}). While intra-BS interference $\varphi_k^\text{intra}$ is directly controllable by the device's serving BS, the inter-BS interference $\varphi_k^\text{inter}$ originates from external sources and cannot be directly managed by the local BS.
For illustration purposes, Fig.~\ref{fig:channel_knowledge} shows an example of a network that includes two BSs, five XR devices, and BS-device links. These links are categorized as communication links (solid lines) or inter-BS interference links (dashed lines). For instance, based on Fig.~\ref{fig:channel_knowledge}, we observe the perspective of BS $1$ performing resource allocation for its associated devices. From BS $1$'s perspective, known channels are depicted in green, while unknown channels are shown in red. For example, BS $1$ estimates its associated intra-BS channels $\bm{\hat{h}}_{1,1}$, $\bm{\hat{h}}_{1,2}$, and $\bm{\hat{h}}_{1,3}$, as well as the inter-BS interfering channels $\bm{\hat{h}}_{1,4}$ and $\bm{\hat{h}}_{1,5}$ (i.e., the channels connecting BS $1$ to XR devices $4$ and $5$), through uplink pilot measurements in time-division duplex mode or via user feedback in frequency-division duplex mode \cite{8255833}. BS $1$ is, therefore, able to compute $\varphi_4^\text{inter}$ and $\varphi_5^\text{inter}$ via \eqref{eq:phi_inter}. Similarly, BS $2$ computes $\varphi_1^\text{inter}$, $\varphi_2^\text{inter}$, and $\varphi_3^\text{inter}$. Thus, ideally, each BS $b$ informs all other BS of its induced inter-BS interference $\varphi_k^\text{inter}$, $\forall k\in\mathcal{K}\backslash\mathcal{K}_b$, based on its locally estimated interfering channels. In networks with more than two BSs, each BS estimates its own set of cross-BS interference channel estimates, and then computes and transmits only its own portion of $\varphi_k^\text{inter}$. The CP, acting as a central parameter server, is then responsible for collecting all inter-BS interference terms, aggregating them, and distributing the aggregated results among all BSs.

A data-driven approach is adopted to address the inter-BS interference management. Exchanging inter-BS interference information for every channel estimation and each uncertainty sample would be computationally expensive and practically infeasible. Instead, our paper makes use of the mean inter-BS interference, referred to as $\bar{\varphi}_{k}$, which is computed and exchanged, aligning with the modular learning framework. This mean interference value can be shared at the end of each learning epoch during training. During inference, the mean inter-BS interference is exchanged initially and requires updates only when channel statistics exhibit significant changes.\vspace*{-.25cm}

\subsection{Decentralized DNN Architecture}
To best empower the distributed FL implementation of the solution framework of problem \eqref{eq:Opt2}, each BS hosts a local DNN to determine its associated devices' resource allocation based on local channel estimates, computation requirement estimates, and mean inter-BS interference levels, as depicted in Fig.~\ref{fig:dnn_mdl}. At every BS $b$, the input layer consists of $2L_bK_b+2K_b$ and the output layer of $2K_b+1$ neurons. The DNN then returns $\bm{x}_b=[x_1,\cdots,x_{2K_b+1}]^T$. Compared to the centralized architecture in Fig.~\ref{fig:dnn_mdl}, the distributed DNN (with modifications highlighted in green) considers only $K_b$ users and the input layer is extended by additional $K_b$ neurons that represent the mean inter-BS interference $\bar{\varphi}_{k}$. 
We let $\mathcal{F}_{b,\boldsymbol{\Theta}_b}$ denote the proposed DNN for BS $b$, such that problem \eqref{eq:Opt2} is solved by\vspace*{-.1cm}
\begin{align}
\bm{x}_b = \mathcal{F}_{b,\boldsymbol{\Theta}_b}(\bm{\hat{h}}_b,\bm{\hat{\omega}}_b,\bar{\boldsymbol{\varphi}}), \label{eq:dnnFL}\\[-.7cm]\nonumber
\end{align}
where $\bm{\hat{h}}_b=[\bm{\hat{h}}_{b,1}^T,\cdots,\bm{\hat{h}}_{b,K}^T]^T$ and $\boldsymbol{\hat{\omega}}_b=[\hat{\omega}_{1},\cdots,\hat{\omega}_{K_B}]^T$. Note that $\boldsymbol{\Theta}_b$ refers to the $b$'th BS's DNN model parameters. We map the DNN output $\bm{x}_b$ to $\bm{p}_b^\text{tx}$ and $p_b^\text{co}$ using equation \eqref{eq:ptx_pco} and to $\bm{f}_b^\text{pow}$ using equation \eqref{eq:fco}. Afterwards, we inject both channel and computing state information error samples to compute the empirical $\gamma$-th quantile worst-case delay $\hat{t}_{b,\text{max}}^{\hspace*{.02cm}\gamma}$ similar to the centralized solution.\vspace*{-.45cm}

\subsection{Federated Deep Learning for Robust Delay Optimization}\vspace*{-.05cm}
Fundamentally, distributed optimization is best enabled in a network of multiple entities, e.g., BSs, that solve a global optimization problem, such as the one in \eqref{eq:Opt2}, by using localized computations collectively \cite{10636212}. 
Specifically, FL describes a distributed optimization methodology for collaborative AI model training. In the context of our paper, we consider the $B$ BSs as clients, each hosting a local AI model for performing the communication and computation resource allocation for their associated XR devices. The BSs are connected to the CP, which hosts a global parameter server responsible for model aggregation, as can be seen in Fig.~\ref{fig:sys_mdl}. 

Our paper, therefore, solves problem \eqref{eq:Opt2} using a federated deep learning solution, which partially builds upon the steps of FedAvg-Algorithm \cite{mcmahan2017communication}\footnote{Note that the original steps according to the FedAvg-Algorithm are marked in italic font in Algorithm~\ref{alg:FLtrain} description.}.
More precisely, initially, each BS performs local model updates using the gradient $\partial\hat{t}_{b,\text{max}}^{\hspace*{.02cm}\gamma}/\partial\Theta_b$, based on their local channel and computing requirement estimations. Further, each BS computes its portion of the inter-BS interference $\varphi_k^\text{inter}$, $\forall k\in\mathcal{K}\backslash\mathcal{K}_b$ using \eqref{eq:varphik}. Then all BSs upload their local model parameters $\bm{\Theta}_b$ and inter-BS interference $\varphi_k^\text{inter}$, $\forall k\in\mathcal{K}\backslash\mathcal{K}_b$ to the CP. The CP aggregates the models to form the global model $\bm{\Theta}^\text{global}$ by taking the average of $\bm{\Theta}_b$, $\forall b\in\mathcal{B}$ as follows\vspace*{-.2cm}
\begin{equation}
\bm{\Theta}^\text{global} = \frac{1}{B} \sideset{}{_{b\in\mathcal{B}}}{\sum} \bm{\Theta}_b. \label{eq:model_aggr}\\[-.15cm]
\end{equation}
Afterwards, the CP combines all inter-BS interference terms to obtain $\varphi_k^\text{inter}$, $\forall k\in\mathcal{K}$. In the last step, the CP broadcasts the global model parameters $\bm{\Theta}^\text{global}$ and the inter-BS interference $\varphi_k^\text{inter}$, $\forall k\in\mathcal{K}$ to all BSs. The whole process is repeated until a convergence criterion is met or the predefined number of training epochs is reached. The above steps are summarized in the description of Algorithm~\ref{alg:FLtrain} below.\vspace*{-.25cm}

\begin{algorithm}[h]
\caption{Proposed FL-based Training Procedure}
\begin{algorithmic}[1]
	\REPEAT
	\FORALL{BSs $b\in\mathcal{B}$}
	\STATE \textit{Local model update} using gradient $\partial\hat{t}_{b,\text{max}}^{\hspace*{.02cm}\gamma}/\partial\Theta_b$
	\STATE Compute portion of inter-BS interference $\varphi_k^\text{inter}$, $\forall k\in\mathcal{K}\backslash\mathcal{K}_b$ using \eqref{eq:phi_inter}
	\STATE Upload local model parameters $\bm{\Theta}_b$ and inter-BS interference $\varphi_k^\text{inter}$, $\forall k\in\mathcal{K}\backslash\mathcal{K}_b$ to CP
	\ENDFOR
	\STATE Model \textit{aggregation} at CP to form global model $\bm{\Theta}^\text{global}$ by taking average of $\bm{\Theta}_b$, $\forall b\in\mathcal{B}$, via \eqref{eq:model_aggr}
	\STATE CP adds inter-BS interferences to obtain $\varphi_k^\text{inter}$, $\forall k\in\mathcal{K}$%
	\STATE \textit{Broadcasting} global model $\bm{\Theta}^\text{global}$ and inter-BS interference $\varphi_k^\text{inter}$, $\forall k\in\mathcal{K}$, to all BSs
	\UNTIL{Convergence}
\end{algorithmic}
\label{alg:FLtrain}
\end{algorithm}%
Note that the FedAvg algorithm, the steps of which are partially used in Algorithm~\ref{alg:FLtrain}, may not guarantee convergence under highly heterogeneous data distributions \cite{li2019convergence}. While a formal convergence analysis is beyond the scope of this paper, the simulations in Sec.~\ref{ssec:conv} empirically demonstrate convergence within a practical number of epochs under uncertainty injection.\looseness-1
\vspace*{-.4cm}

\subsection{Data Exchange}\vspace*{-.1cm}
This subsection analyzes the data exchange requirements across the system nodes, both for the proposed centralized (i.e., Algorithm~\ref{alg:CENtrain}) and for the proposed FL (i.e., Algorithm~\ref{alg:FLtrain}) approaches. In the centralized learning setup, a CP manages all BSs. In contrast, the FL setup relies on a common parameter server, while the BSs perform local operations independently.

The centralized learning approach requires all BSs $b\in\mathcal{B}$ to upload each devices' estimated channel coefficients $\bm{\hat{h}}_{b,l,k}$, $\forall (l,k)\in(\mathcal{L}_b,\mathcal{K})$ and estimated computing requirements $\hat{\omega}_{b,k}$, $\forall k\in\mathcal{K}$ to the CP. In practical operations, each BS sends the channel coefficients and computation requirements for each new acquired channel or computing requirement estimation. Further, the CP broadcasts $\bm{p}^\text{tx}$, $\bm{p}^\text{co}$, and $\bm{f}^\text{co}$. The learning procedure requires a number of $E$ epoches, with $M$ minibatches each, with a size of $\nu_m$, and a validation size of $\nu_v$. {Additionally, only during training, for each channel and computing estimation, the BSs send $N$ sampled channel and computing uncertainties to the CP.} The channel coefficients are transmitted as complex floats with size $D_\text{cfloat}=64$ bits, while the computing requirements, power loadings, and computing allocations are transmitted as floats with size $D_\text{float}=32$ bits. The overall data exchange for centralized training becomes\vspace*{-.1cm}
\begin{align}
\Omega^{\text{cen}}\!=\!E N\!\left(M \nu_m+\nu_v\right)\left((3K+B) D_\text{float}\!+\!L K D_\text{cfloat}\right).\label{eq:omega_cen}\\[-.65cm]\nonumber
\end{align}
The formulation \eqref{eq:omega_cen} captures the total data exchange required for centralized learning across multiple epochs and minibatches, factoring in parameter data sizes, the number of BSs and antennas, and the number of devices.

The FL approach facilitates local model updates at the BSs and the exchanged data originates from the $i)$ broadcasting and $iii)$ aggregation step. In each communication round, the CP broadcasts the most recent global model $\mathcal{F}_{\boldsymbol{\Theta}}(\cdot)$, with parameters ${\boldsymbol{\Theta}}$, and the average per-device inter-BS interference $\bar{\varphi}_k$, $\forall k\in\mathcal{K}$. In the aggregation step, each BS uploads its local DNN $\mathcal{F}_{\boldsymbol{\Theta}_b}^\text{local}(\cdot)$, with parameters ${\boldsymbol{\Theta}}_b$, and the locally-computed average inter-BS interference $\bar{\varphi}_{k'}$ to all $k'\in\{\mathcal{K}\backslash\mathcal{K}_b\}$. The DNN has a total of $P$ parameters, consisting of $\eta_h$ hidden layers, with $n_h$ neurons each, $n_\text{in}$ input layer neurons, and $n_\text{out}$ output layer neurons. The DNN parameter size consisting of weights and biases can thus be calculated as\vspace*{-.15cm}
\begin{align}
P &= n_\text{in} n_h + (\eta_h-1) n_h^2 + n_\text{out} n_h + \eta_h n_h + n_\text{out},\\[-.65cm]\nonumber
\end{align}
where the input and output layers consist of $n_\text{in}=2K_b^2+2K_b$ and $n_\text{out}=2K_b+1$ neurons, respectively. Using $P$, the overall data exchange in FL, including both broadcasting and aggregation phases, becomes\vspace*{-.15cm}
\begin{align}
\Omega^{\text{fed}} &= E\big( (P D_\text{float}+KD_\text{float}) \nonumber\\
&\qquad + B(P D_\text{float}+(B-1)K_bD_\text{float}) \big).\label{eq:omega_fed}\\[-.7cm]\nonumber
\end{align}
A numerical comparison of data exchange requirements for both learning paradigms (i.e., Algorithm~\ref{alg:CENtrain} and Algorithm~\ref{alg:FLtrain}) is further presented in Section~\ref{ssec:dat_ex}. 

{We note that uplink communication between devices and BSs (e.g., for channel estimation) is an inherent part of the underlying XR network. The communication overhead analysis above, instead, focuses on the additional overhead introduced by the learning framework, namely model parameter exchange and inter-BS information sharing between BSs and CP, which represent the dominant incremental communication cost.}\vspace*{-.2cm}

\section{Simulation Results and Discussion}\label{sec:sim}	
\begin{table}[t]
\vspace*{.15cm}
\renewcommand{\arraystretch}{1.1}
\centering
\begin{tabular}{l l l}
	\hline
	\hspace*{-.1cm}{\textbf{General}}\hspace*{-.1cm} & {\textbf{Communication}}\hspace*{-.3cm} & {\textbf{Computation}}\\\hline
	\hspace*{-.1cm}$P^\text{max}=34$ dBm\hspace*{-.1cm} & $W=10$ MHz\hspace*{-.3cm} & $F^\text{max}=4.6\cdot 10^9$ cycles/s \\
	\hspace*{-.1cm}$N=1000$ samples\hspace*{-.1cm} & $D_k^\text{out}=7.5\hspace{-.05cm}\cdot\hspace{-.05cm} 10^4$ bits\hspace*{-.3cm} & $D_k^\text{in}=5\cdot 10^4$ bits\\
	\hspace*{-.1cm}$\gamma=0.05$\hspace*{-.1cm} & $\alpha=0.2$ RZF factor\hspace*{-.3cm} & $\tau = 10^{-28}$, $\mu = 3$\\
	\hspace*{-.1cm}  \hspace*{-.1cm} &  $\sigma^2 = -175$ dBm/Hz \hspace*{-.3cm} &  $\omega_{k}\sim \Gamma(2,200)$, $\forall k\in\mathcal{K}$ \\\hline
\end{tabular}
\vspace*{-.15cm}
\caption{Simulation parameters (unless otherwise mentioned).}
\label{tb:simparam}
\vspace{-.6cm}
\end{table}%
In the following, we numerically evaluate the performance of the proposed FL-based solution with joint uncertainty injection. The network consists of $B=3$ BS with $L_b=6$ antennas, $\forall b\in\mathcal{B}$. Unless otherwise mentioned, we place the three BS on a two dimensional plane at $((0.25,0.25),(1.0,0.25),(0.625,0.625))$ given in km with a height of $0.1$ km. The devices are placed randomly around each BS with a radius of $0.25$ km. The default number of XR devices is $K_b=5$, $\forall b\in\mathcal{B}$. 
Other parameters are summarized in Tab.~\ref{tb:simparam}. 
We model the channel coefficients as
\begin{align}
h_{b,l,k} = c\,(4\pi d_{b,k} f_c)^{-1} F_{b,l,k},
\end{align}
with $c$ denoting the speed of light, $d_{b,k}$ denoting the distance between BS $b$ and device $k$ in meter, $f_c = 3$ GHz denoting the carrier frequency, and $F_{b,l,k}$ denoting the fading distribution \cite{9080558}. We assume Rayleigh fading $F_{b,l,k}\sim \mathcal{CN}(0,1)$, $\forall (b,l,k)\in(\mathcal{B},\mathcal{L},\mathcal{K})$, utilize minimum mean-square estimation for the estimated channels $\bm{\hat{h}}$, and set the beamforming vectors $\bm{v}_k$, $\forall k\in\mathcal{K}$, according to a regularized zero-forcing (RZF) beamforming strategy \cite{10071987}, fixing the beamforming vectors $\bm{v}_k$ using: $\bm{V}=\bm{\hat{H}}(\bm{\hat{H}}\bm{\hat{H}}^H+\alpha\bm{I})^{-1}$, where $\bm{V}=[\bm{v}_1,\cdots,\bm{v}_K]$, $\bm{\hat{H}}=[\bm{\hat{h}}_1,\cdots,\bm{\hat{h}}_K]$, $\alpha$ is the RZF factor, and $\bm{I}$ is a $K\times K$ identity matrix. Each $\bm{v}_k$ is normalized to unit norm.
As the beamforming vectors are determined a priori, we modify each DNN's input layer, and only input effective channels, i.e., $\hat{h}_{i,j}^\text{eff}=\bm{\hat{h}}_i^H \bm{v}_j$, $\forall i,j\in\mathcal{K}$, thereby reducing the input layer to $2K^2+K$ nodes for Algorithm~\ref{alg:CENtrain} and $2K_b^2+2K_b$ nodes for Algorithm~\ref{alg:FLtrain}.
The ground-truth computation intensities are drawn according to a Gamma distribution \cite{10007803}. For the DNN parameters and training procedure, we use $\eta_h = 10$ fully-connected hidden layers with $n_h = 400$ neurons each for Algorithm~\ref{alg:CENtrain}, and $\eta_h = 7$ with $n_h = 200$ for Algorithm~\ref{alg:FLtrain}. We apply ReLU activation and train for $E=500$ epochs using $M=50$ minibatches, each containing $\nu_m=1000$ independent network realizations.
Validation and test sizes are set to $\nu_v=2000$ each, where each realization experiences uncertainty injection to obtain the robust objectives. 
We note that the adopted training procedure is implemented in an offline fashion. Such a procedure is in essence based on the channel and computing statistics, which are relatively slow-varying parameters as compared to the overall algorithmic training time. Such a characteristic of the proposed approach lessens the need for frequent training operations.

\subsubsection{Benchmark Schemes}
To benchmark the proposed robust FL-based scheme (referred to as \emph{Robust FL}), i.e., Algorithm~\ref{alg:FLtrain}, we compare it against a regular deep learning scheme (\emph{Non-robust FL}), which is trained without uncertainty injection. Based on Algorithm~\ref{alg:CENtrain}, we also consider centralized worst-case delay minimization schemes as upper bounds to the decentralized learning approach, namely, a robust (\emph{Robust Cen}) and a non-robust (\emph{Non-robust Cen}) variant. 

Additionally, we adopt three optimization-based schemes as benchmarks to best assess the performance of the proposed AI-native (data-based) algorithms. More specifically, the non-robust version of problem \eqref{eq:Opt1}, assuming perfect channel and computation knowledge, is solved using fractional programming (FP) by the benchmark scheme named \emph{Non-robust FP} \cite{8314727}. {The \emph{Uniform FP} scheme equally distributes transmit power and computing resources across devices, optimizing only the ratio of total computing to transmit power via FP \cite{8314727}.} Finally, the \emph{SAA/WMMSE} scheme combines SAA and WMMSE to solve an adapted version of problem \eqref{eq:Opt1}, where the quantile constraint \eqref{eq:t_percentile} is replaced by sample averages \cite{reifert2021rate}.

In Section~\ref{ssec:jui}, we benchmark joint uncertainty injection against two modified schemes: \emph{Comm UI FL}, which accounts only for channel uncertainty, and \emph{Comp UI FL}, which accounts only for computing uncertainty. {Similarly, in Section~\ref{ssec:noniid}, we analyze the performance under heterogeneous uncertainty scenarios and benchmark our proposed schemes against \emph{FedProx} \cite{MLSYS2020} and a local learning heuristic (hereafter denoted by \emph{Local}).}

For the sake of completeness, the detailed description of the above benchmarking optimization problems can be found in the appendix of the paper.

\subsubsection{Practical Channel Estimation Error}
Based on the model in \cite{6409502}, we assume block fading channels, with the XR devices carrying out channel estimation by averaging over $N$ noisy pilot symbols. Furthermore, the power allocated for pilot transmission is not the same as for the actual transmission. The relation of both power allocations can be written as $p_s = \beta p^{(1-\gamma)}$, with $p_s$ being the pilot symbol transmit power, where $\gamma$ and $\beta$ are system parameters. As can be seen in \cite{10361526}, the channel estimation error variance $\sigma_h^2$ is a function of such system parameters given by
\begin{align}
\sigma_h^2 = {p^{-4(\gamma+1)}\big(2\beta Np^{2(\gamma+1)}+1\big)}{(\beta^2 N^2)^{-1}}.
\end{align}
{With $\gamma=0.2$, $\beta=0.4$, and $N=L$, the resulting channel estimation error variance $\sigma_h^2$ spans the range shown in Fig.~\ref{fig:y_delay_x_commV} by moderate variations of the underlying system parameters (e.g., pilot length, transmit power, and noise level). This shows how the considered values of $\sigma_h^2$ correspond to realistic operating regimes rather than isolated parameter choices.}

\subsubsection{Practical Computing Estimation Error}
{The computation uncertainty parameter $\sigma_\omega^2$ models variability in processing requirements due to dynamic XR workloads, such as field-of-view rendering and encoding. Since precise workload statistics depend on the application and system implementation \cite{lorch2001improving}, we consider a range of $\sigma_\omega^2$ values in the simulations (see Fig.~\ref{fig:y_delay_x_compV}). The value $\sigma_\omega^2=3200$ is selected as a representative intermediate point, while the overall conclusions are drawn based on trends observed across different uncertainty levels.}

\subsection{Impact of Communication and Computation Uncertainty}
\begin{figure}[t]
\centering
\includegraphics[width=.9\linewidth]{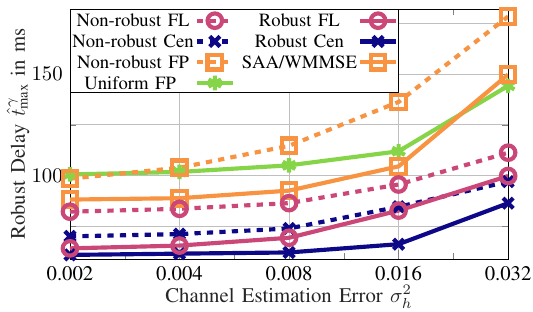}\\
\vspace*{-.3cm}
\caption{Robust delay $\hat{t}_\text{max}^{\hspace*{.02cm}\gamma}$ over channel uncertainty, i.e., $\sigma_h^2$.}
\label{fig:y_delay_x_commV}
\vspace*{-.55cm}
\end{figure}
In the first set of simulations, we analyze the impact of channel and computing requirement uncertainties on the robust delay, $\hat{t}_\text{max}^{\hspace*{.02cm}\gamma}$. 
To this end, Fig.~\ref{fig:y_delay_x_commV} illustrates the robust delay as a function of the channel estimation error variance $\sigma_h^2$, while keeping the computation estimation error variance fixed at $\sigma_\omega^2 = 3200$. 
As a first observation, ML-based schemes (i.e., \emph{Non-robust FL}, \emph{Non-robust Cen}, \emph{Robust FL}, and \emph{Robust Cen}) outperform the baselines across all tested values, achieving delay reductions of $6$ to $56$ ms. Higher values of the channel uncertainty (e.g., in dynamic environments) even widens the performance gap between all ML-optimized schemes and the baselines.
Fig.~\ref{fig:y_delay_x_commV} shows that up to $\sigma_h^2=0.016$, \emph{SAA/WMMSE} is the best-performing optimization baseline, achieving slightly lower $\hat{t}_\text{max}^{\hspace*{.02cm}\gamma}$ than \emph{Non-robust FL}. However, since \emph{SAA/WMMSE}, as also illustrated in \cite{reifert2021rate}, relies on SAA, it provides robustness only in terms of the average performance across samples, which explains its inability to match the robust ML-based schemes.
Among the illustrated ML-based methods is the fully centralized approach (\emph{Robust Cen}) that consistently achieves the lowest robust delay. While \emph{Robust FL} performs similarly at low $\sigma_h^2$, the gap between the two schemes increases as channel uncertainty grows. Despite not matching Algorithm~\ref{alg:CENtrain}'s performance, the \emph{Robust FL} (i.e., the scheme associated with Algorithm~\ref{alg:FLtrain}) offers a practical alternative with robust delay values close to the centralized scheme.
Importantly, Fig.~\ref{fig:y_delay_x_commV} highlights the advantage of robust (uncertainty-aware) learning: Both robust centralized and FL schemes significantly outperform their non-robust counterparts (i.e., \emph{Non-robust FL} and \emph{Non-robust Cen}). For instance, at $\sigma_h^2=0.008$, \emph{Robust FL} and \emph{Robust Cen} achieve $18$ ms and $12$ ms lower robust delay, respectively, than their non-robust counterparts. 
Interestingly, uncertainty injection has a more pronounced impact on FL-based schemes, emphasizing its importance in distributed architectures. 
These findings underscore three significant insights: Uncertainty injection is crucial for minimizing robust delays in general, the proposed FL-based scheme performs closely to its centralized counterpart, and the joint uncertainty injection scheme is effective across varying uncertainty scenarios.

\begin{figure}[t]
\centering
\includegraphics[width=.9\linewidth]{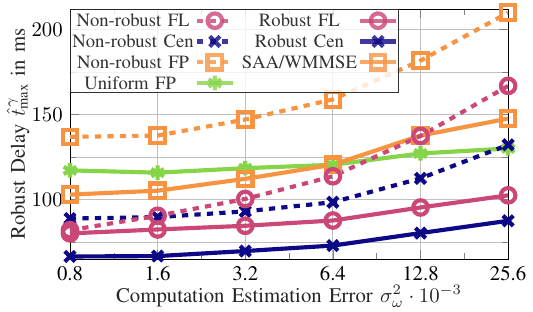}\\
\vspace*{-.3cm}
\caption{Robust delay $\hat{t}_\text{max}^{\hspace*{.02cm}\gamma}$ over computing uncertainty, i.e., $\sigma_\omega^2$.}
\label{fig:y_delay_x_compV}
\vspace*{-.55cm}
\end{figure}
Fig.~\ref{fig:y_delay_x_compV} reaffirms the advantages of Algorithm~\ref{alg:FLtrain} and uncertainty injection by depicting the robust delay as a function of the computation estimation error variance $\sigma_\omega^2$, with fixed $\sigma_h^2 = 0.02$.
Fig.~\ref{fig:y_delay_x_compV} readily depicts how \emph{Uniform FP} consistently performs worse than all ML-based schemes at low $\sigma_\omega^2$. However, as computation uncertainty increases, non-robust ML-based schemes (\emph{Non-robust FL} and \emph{Non-robust Cen}) become increasingly sensitive, ultimately losing their advantage over the uniform power allocation scheme at $\sigma_\omega^2 = 6400$ for \emph{Non-robust FL} and $\sigma_\omega^2 = 12800$ for \emph{Non-robust Cen}. Similarly, \emph{Non-robust FL} also loses its advantage over \emph{SAA/WMMSE} at $\sigma_\omega^2 = 12800$. This highlights that ML-based optimization without explicit uncertainty awareness does not guarantee robust performance.
A deeper comparison with Fig.~\ref{fig:y_delay_x_commV} reveals an interesting remark. While \emph{Robust FL} is more severely affected by communication uncertainty, Fig.~\ref{fig:y_delay_x_compV} shows that computation uncertainty has a more moderate impact. \emph{Robust FL} maintains a consistent $14$ ms gap from \emph{Robust Cen}, demonstrating a relatively smaller degradation in high-computation-uncertainty regimes. Such an observation is in fact reasonable, as distributed algorithms like Algorithm~\ref{alg:FLtrain} are inherently more sensitive to communication uncertainties, affecting the interference levels but not the computation process.\looseness-1

Overall, both robust uncertainty injection schemes (\emph{Robust FL} and \emph{Robust Cen}) improve the network-wide robust delay performance, albeit depicting varying effectiveness across uncertainty conditions. Algorithm~\ref{alg:FLtrain}, particularly, maintains a stable performance gap toward the optimal centralized solution, outperforms non-robust baselines, and demonstrates strong practicality for real-world implementation.\vspace*{-.2cm} 

\subsection{Advantages of Joint Uncertainty Injection}\label{ssec:jui}
\begin{figure}[t]
\centering
\includegraphics[width=.9\linewidth]{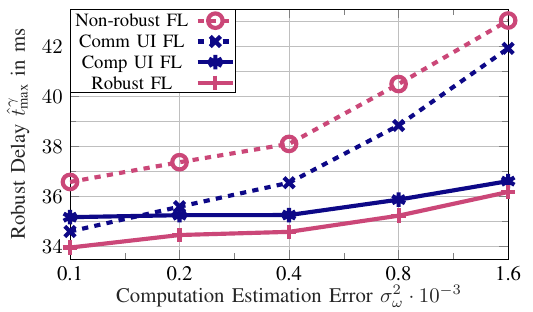}\\
\vspace*{-.3cm}
\caption{Robust delay $\hat{t}_\text{max}^{\hspace*{.02cm}\gamma}$ over computing uncertainty, i.e., $\sigma_\omega^2$.}
\label{fig:y_delay_x_compV_singleUI}
\vspace*{-.55cm}
\end{figure}
In the next set of simulations, we analyze the impact of two different uncertainty injection sub-schemes: \emph{Comm UI FL}, which injects only channel uncertainty during training, and \emph{Comp UI FL}, which injects only computing uncertainty. Fig.~\ref{fig:y_delay_x_compV_singleUI} depicts the robust delay $\hat{t}_\text{max}^{\hspace*{.02cm}\gamma}$ as a function of the computation estimation error variance $\sigma_\omega^2$, with $\sigma_h^2 = 0.01$.
It is first worthy to note that all uncertainty injection schemes outperform the non-robust baseline implementation (\emph{Non-robust FL}), achieving delay reductions between $1$ and $7$ ms. Notably, Algorithm~\ref{alg:FLtrain} (\emph{Robust FL}), which leverages joint uncertainty injection, consistently achieves the lowest robust delay across all levels of uncertainty.
Fig.~\ref{fig:y_delay_x_compV_singleUI} further highlights the adaptability of Algorithm~\ref{alg:FLtrain} across different uncertainty scenarios. For lower values of $\sigma_\omega^2$, where communication uncertainty dominates, \emph{Comm UI FL} performs comparably to \emph{Robust FL}, whereas \emph{Comp UI FL} is less effective. Conversely, at higher computing uncertainties (e.g., $\sigma_\omega^2 = 800$), \emph{Comp UI FL} and \emph{Robust FL} yield similar performances, while \emph{Comm UI FL} suffers from significant delay increases.
%
These findings underscore the effectiveness of joint uncertainty injection in improving network-wide robust delay performance, with varying effectiveness depending on the uncertainty type. Notably, Algorithm~\ref{alg:FLtrain} not only generalizes both \emph{Comp UI FL} and \emph{Comm UI FL} but also consistently achieves the lowest robust delays, demonstrating its flexibility and superiority across diverse uncertainty conditions.\vspace*{-.2cm}

\subsection{Scalability and Density}
\begin{figure}[t]
\centering
\includegraphics[width=.9\linewidth]{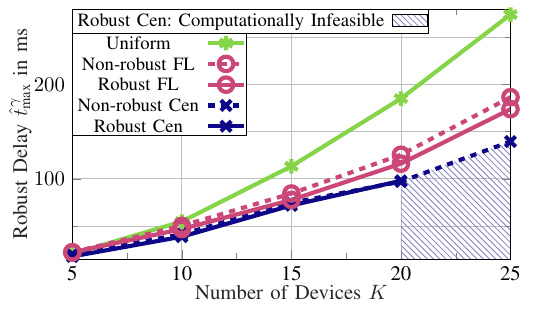}\\
\vspace*{-.25cm}
\caption{Robust worst-case delay $\hat{t}_\text{max}^{\hspace*{.02cm}\gamma}$ vs. number of devices $K$.}
\label{fig:y_delay_x_numUE}
\vspace*{-.55cm}
\end{figure}
In this simulation set, we evaluate the algorithmic scalability by fixing the channel and computation estimation error variances to $\sigma_h^2 = 0.016$ and $\sigma_\omega^2 = 3200$, respectively. The scenario considers an expanded network with $B = 5$ BSs, positioned at $((0.25,0.25),(1,0.25),(0.625,0.625),(0.25,1.25),(1,1.25))$ in km, each equipped with $L_b = 4$ antennas. The number of devices $K$ is varied from $5$ to $25$, allowing us to analyze the scalability of the proposed learning-based approaches in dense networks. Fig.~\ref{fig:y_delay_x_numUE} illustrates the robust delay as a function of $K$, comparing the proposed FL-based and centralized uncertainty injection schemes against the uniform baseline.
The results reveal that, as the number of devices increases, the proposed ML-based methods become increasingly more effective as compared to the uniform allocation scheme. For instance, at $K = 20$, \emph{Robust FL} achieves a robust delay reduction of $69$ ms relative to the uniform power allocation (\emph{Uniform}), while the robust centralized scheme (\emph{Robust Cen}) reduces the delay by $88$ ms. However, at high device densities, centralized solutions gain a larger advantage over FL-based approaches. The gap between \emph{Robust FL} and \emph{Robust Cen} widens with growing $K$, as the impact of the emerging interference in multi-device networks becomes more pronounced. At $K = 15$, the gap between the two schemes is $6$ ms, which increases to $19$ ms at $K = 20$.
However, the computational complexity of Algorithm~\ref{alg:CENtrain} grows significantly, to the extent that for $K=25$, the centralized solution, i.e., \emph{Robust Cen}, becomes unable to determine a reliable solution within a reasonable time. This computational infeasibility region is marked in a shaded area in Fig.~\ref{fig:y_delay_x_numUE}.
When comparing uncertainty injection against non-robust ML-based resource allocation, \emph{Robust FL} consistently outperforms \emph{Non-robust FL}, with the performance gap widening as the number of devices increases. While both schemes perform similarly at $K = 5$, the delay gap increases to $7$ ms at $K = 15$ and further to $13$ ms at $K = 25$. 
These results underscore three key findings. First, ML-based resource allocation offers substantial performance gains in XR networks, especially at large numbers of XR devices. Second, while FL-based methods provide decentralized and scalable solutions, centralized learning remains superior in minimizing delay under high device densities. Finally, joint uncertainty-aware training plays a crucial role in maintaining robust performance, particularly in dynamic and interference-prone network environments.

\begin{figure}[t]
\centering
\includegraphics[width=.9\linewidth]{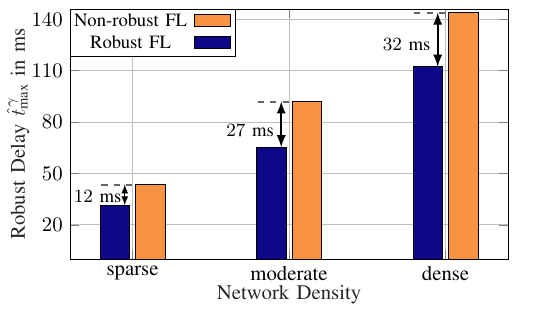}\\
\vspace*{-.3cm}
\caption{Robust delay $\hat{t}_\text{max}^{\hspace*{.02cm}\gamma}$ over network density.}
\label{fig:y_delay_x_numUEdensity_bar}
\vspace*{-.5cm}
\end{figure}
\begin{table}[t]
\vspace*{.2cm}
\renewcommand{\arraystretch}{1.2}
\centering
\begin{tabular}{r c c c c c c}
	\hline 
	\hspace*{-.2cm}{\textbf{Network density}}\hspace*{-.2cm}& 
	\multicolumn{2}{c}{{\textbf{sparse}}\hspace*{-.2cm}}& \multicolumn{2}{c}{{\textbf{moderate}}\hspace*{-.2cm}}& \multicolumn{2}{c}{{\textbf{dense}}}\\
	{\textbf{Unit in ms}}\hspace*{-.2cm}& 
	{$\boldsymbol{t^\text{dnn}_\text{inf}}$}\hspace*{-.25cm}&
	{$\boldsymbol{\hat{t}_\text{max}^{\hspace*{.02cm}\gamma}}$}\hspace*{-.2cm}&
	{$\boldsymbol{t^\text{dnn}_\text{inf}}$}\hspace*{-.25cm}&
	{$\boldsymbol{\hat{t}_\text{max}^{\hspace*{.02cm}\gamma}}$}\hspace*{-.2cm}&
	{$\boldsymbol{t^\text{dnn}_\text{inf}}$}\hspace*{-.25cm}&
	{$\boldsymbol{\hat{t}_\text{max}^{\hspace*{.02cm}\gamma}}$}\\\hline
	Non-robust FL \hspace*{-.2cm}
	& 4.498 \hspace*{-.25cm}& 43.717\hspace*{-.2cm}
	& 4.301\hspace*{-.25cm}& 91.918\hspace*{-.2cm}
	& 4.525\hspace*{-.25cm}& 144.651\hspace*{-.2cm}\\
	Robust FL \hspace*{-.2cm}
	& 4.408 \hspace*{-.25cm}& 31.392\hspace*{-.2cm}
	& 4.199\hspace*{-.25cm}& 65.398\hspace*{-.2cm}
	& 4.288\hspace*{-.25cm}& 112.939\hspace*{-.2cm}\\
	Non-robust FP \hspace*{-.2cm}
	&8.5{\scriptsize$\cdot10^{3}$}\hspace*{-.25cm}& 59.847\hspace*{-.2cm}
	&12.2{\scriptsize$\cdot10^{3}$}\hspace*{-.25cm}& 129.592\hspace*{-.2cm}
	&20.2{\scriptsize$\cdot10^{3}$}\hspace*{-.25cm}& 212.580\hspace*{-.2cm}\\
	SAA/WMMSE \hspace*{-.2cm}
	&9.6{\scriptsize$\cdot10^{3}$}\hspace*{-.25cm}& 42.520\hspace*{-.2cm}
	&38.6{\scriptsize$\cdot10^{3}$}\hspace*{-.25cm}& 96.609\hspace*{-.2cm}
	&104.6{\scriptsize$\cdot10^{3}$}\hspace*{-.25cm}& 176.476\hspace*{-.2cm}\\
	Uniform FP \hspace*{-.2cm}
	&6.4{\scriptsize$\cdot10^{3}$}\hspace*{-.25cm}& 42.582\hspace*{-.2cm}
	&9.9{\scriptsize$\cdot10^{3}$}\hspace*{-.25cm}& 100.187\hspace*{-.2cm}
	&16.9{\scriptsize$\cdot10^{3}$}\hspace*{-.25cm}& 178.833\hspace*{-.2cm}\\\hline
\end{tabular}
\vspace*{-.15cm}
\caption{Mean inference/optimization time $\boldsymbol{t^\text{dnn}_\text{inf}}$ for ML-based schemes and optimization baselines.}
\label{tb:runtime}
\vspace{-.5cm}
\end{table}%
Along the above discussion, we analyze the behavior of Algorithm~\ref{alg:FLtrain} in both its robust and non-robust implementations within networks of varying density. To this end, we increase the number of antennas per BS to $L_b = 12$ while fixing the channel and computation estimation error variances at $\sigma_h^2 = 0.03$ and $\sigma_\omega^2 = 1600$, respectively.
Fig.~\ref{fig:y_delay_x_numUEdensity_bar} illustrates the robust delay as a function of network density, where the x-axis represents three density levels, i.e., sparse, moderate, and dense, corresponding to $K_b = 4$, $K_b = 8$, and $K_b = 12$ devices per BS. 
The results indicate that increasing network density significantly degrades the robust delay performance. For instance, in a sparse network, \emph{non-robust FL} and \emph{robust FL} achieve delays of $44$ ms and $32$ ms, respectively. In contrast, in a dense network, the robust delay increases to $145$ ms for \emph{non-robust FL} and $113$ ms for \emph{robust FL}.
Tab.~\ref{tb:runtime} additionally shows the mean inference/optimization time (denoted by $t^\text{dnn}_\text{inf}$) alongside the robust delay for both FL-based schemes and optimization baselines for sparse, moderate, and dense networks. Tab.~\ref{tb:runtime} particularly shows how the inference times of \emph{Robust FL} and \emph{Non-robust FL} remain in the order of $4$-$5$ ms, whereas optimization-based schemes require several seconds per channel and computation estimation. Moreover, the inference time of the DNN-based schemes scales sub-linearly, since only the input and output layers are affected by the number of devices.
Interestingly, the gap between \emph{non-robust FL} and \emph{robust FL} increases as the network density grows, rising from $12$ ms in sparse networks to $32$ ms in dense networks. This highlights the increasing importance of joint uncertainty injection, particularly in high-density scenarios, where interference and computational variations are more pronounced.

These results emphasize the growing importance of distributed uncertainty injection in XR-enabled 6G networks, particularly in dense deployments. As the network density increases, the computational complexity of centralized training algorithms becomes prohibitive, necessitating the use of our proposed distributed FL-based mechanisms to ensure scalability and resilience.\vspace*{-.25cm}

\subsection{Heterogeneous Uncertainties}\label{ssec:noniid}
{To model non-IID local data across BSs, each BS is assigned a fixed estimation error variance, while the spread across BSs controls the heterogeneity level. Increasing this spread yields more diverse local uncertainty conditions while preserving the same average uncertainty. We retain the Gaussian error realizations but make their variances BS-dependent, i.e., $h^\text{err}_{b,l,k}\sim\mathcal{CN}(0,\sigma_{h,b}^2)$ and $\omega^\text{err}_{k}\sim\mathcal{N}(0,\sigma_{\omega,b}^2)$, where $\sigma_{h,b}^2 \in [\sigma_h^2-\Delta_h/2, \sigma_h^2+\Delta_h/2]$ and $\sigma_{\omega,b}^2 \in [\sigma_\omega^2-\Delta_\omega/2, \sigma_\omega^2+\Delta_\omega/2]$. For $B=3$, we use evenly spaced values within these intervals. The remaining simulation parameters are set to $\sigma_h^2 = 0.016$, $\sigma_\omega^2 = 3200$, and $E=1000$ training epochs.} 

\begin{figure}[t]
\centering
\includegraphics[width=.9\linewidth]{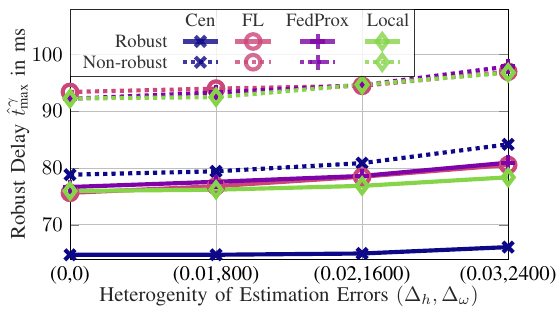}\\
\vspace*{-.25cm}
\caption{Robust worst-case delay $\hat{t}_\text{max}^{\hspace*{.02cm}\gamma}$ vs. data heterogeneity for different FL implementations.}
\label{fig:rev_y_delay_x_nonIIDsigma}
\vspace*{-.55cm}
\end{figure}
{Fig.~\ref{fig:rev_y_delay_x_nonIIDsigma} shows the robust delay versus $(\Delta_h,\Delta_\omega)$. As heterogeneity increases, all schemes experience performance degradation. Under homogeneous uncertainties, i.e., $(\Delta_h,\Delta_\omega)=(0,0)$, \emph{Robust Cen} performs best, followed by \emph{Robust FL}, while \emph{Robust Local} and \emph{Robust FedProx} perform slightly worse. The figure also shows that all non-robust schemes, particularly their federated variants, perform significantly worse.
At high heterogeneity, i.e., $(\Delta_h,\Delta_\omega)=(0.03,2400)$, \emph{Robust Local} outperforms both \emph{Robust FL} and \emph{Robust FedProx} by approximately $2$ ms, while \emph{Robust Cen} still achieves the overall lowest delay. 
These results indicate that \emph{Robust Cen} remains effective under strong uncertainty heterogeneity with only minor performance loss. When the uncertainty levels across BSs are similar, \emph{Robust FL} outperforms both \emph{Robust FedProx} and \emph{Robust Local}, whereas under highly heterogeneous conditions, \emph{Robust Local} becomes advantageous due to its fully independent model adaptation. While \emph{Robust FedProx} does not achieve the best performance in this setup, tuning its proximal parameter $\delta$ allows interpolation between FedAvg ($\delta=0$) and more constrained local updates, which may improve performance under specific heterogeneity regimes. 
Overall, the results show that highly heterogeneous uncertainty scenarios favor either centralized processing or, among distributed approaches, more localized model adaptation.}

\subsection{Impact of Reliability Parameter $\gamma$}\label{ssec:gamma}
\begin{figure}[t]
\centering
\includegraphics[width=.9\linewidth]{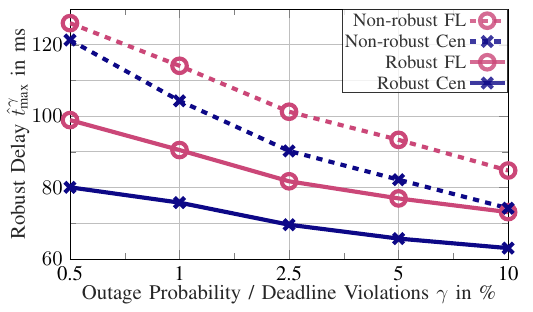}\\
\vspace*{-.25cm}
\caption{Robust worst-case delay $\hat{t}_\text{max}^{\hspace*{.02cm}\gamma}$ vs. parameter $\gamma$, showing impact of hard latency deadline guarantees.}
\label{fig:rev_y_delay_x_gamma}
\vspace*{-.55cm}
\end{figure}
{We investigate the impact of the reliability parameter $\gamma$ in the proposed $\gamma$-th quantile formulation, which directly controls the allowable probability of delay violations. The simulation setup uses $\sigma_h^2 = 0.016$, $\sigma_\omega^2 = 3200$, with training performed over $750$ epochs.
Fig.~\ref{fig:rev_y_delay_x_gamma} illustrates the robust worst-case delay $\hat{t}_\text{max}^{\hspace*{.02cm}\gamma}$ as a function of $\gamma$ for the considered schemes. The results show that increasing $\gamma$ leads to lower delay for all methods, while smaller $\gamma$ values (i.e., stricter reliability requirements) result in higher delay. Furthermore, the performance gap between robust and non-robust schemes becomes more pronounced as $\gamma$ decreases, highlighting the importance of uncertainty-aware designs under stringent latency constraints. 
It is also observed that \emph{Robust Cen} is less sensitive to $\gamma$ and maintains more stable performance, whereas \emph{Robust FL} exhibits a stronger dependence on $\gamma$, leading to a larger performance gap at lower values. Overall, the results demonstrate a fundamental trade-off between delay and reliability, with $\gamma$ serving as a tunable parameter to balance latency performance and deadline violation probability.}

\subsection{Impact of Non-Gaussian Uncertainty Assumptions}\label{ssec:distribution}
\begin{figure}[t]
\centering
\includegraphics[width=.9\linewidth]{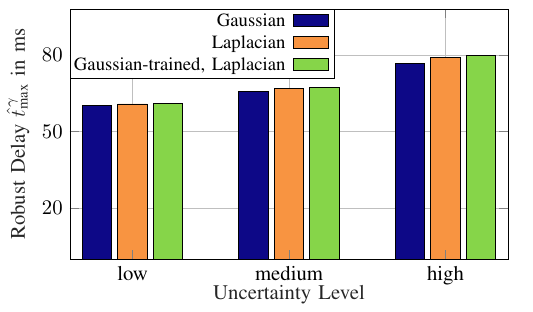}\\
\vspace*{-.25cm}
\caption{Robust worst-case delay $\hat{t}_\text{max}^{\hspace*{.02cm}\gamma}$ vs. three uncertainty levels (\{\emph{low}, \emph{medium}, \emph{high}\} corresponding to $(\sigma_h^2,\sigma_\omega^2)\in\{(0.004,800), (0.008,1600), (0.016,3200)\}$, respectively).}
\label{fig:rev_y_delay_x_uncertainty_bar_LAP}
\vspace*{-.55cm}
\end{figure}
{To further validate the distribution-agnostic nature of the proposed framework, we consider Laplacian uncertainty, which exhibits heavier tails than the Gaussian distribution and captures rare but significant deviations (e.g., abrupt channel variations). Specifically, we model the error as $x \sim \text{Laplace}(0,\sigma_{\text{Laplace}})$ with variance matching the Gaussian case, i.e., $\sigma_{\text{Laplace}}\in\{\sqrt{\sigma_{h}^2 / 2},\sqrt{\sigma_{\omega}^2 / 2}\}$, for either channel or computing uncertainty. Fig.~\ref{fig:rev_y_delay_x_uncertainty_bar_LAP} shows the robust worst-case delay for different uncertainty levels. As expected, increasing uncertainty leads to higher delay. Compared to Gaussian uncertainty, the Laplacian model results in slightly higher delay due to its heavier tails. Importantly, when evaluating a Gaussian-trained model under Laplacian uncertainty, only a minor performance degradation of $0.5$ ms is observed, indicating strong cross-distribution generalization. These results support the distribution-agnostic design of the proposed uncertainty injection framework.}

\subsection{Convergence Behavior}\vspace*{-.0cm}\label{ssec:conv}
\begin{figure}[t]
\centering
\includegraphics[width=.9\linewidth]{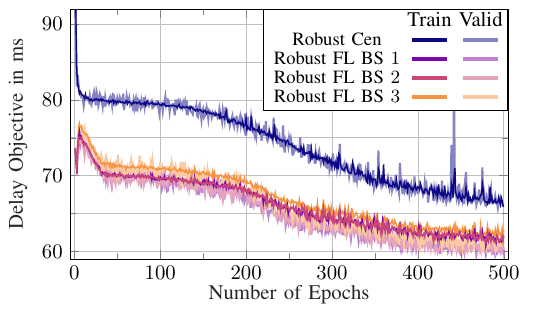}\\
\vspace*{-.3cm}
\caption{Convergence behavior of centralized learning and FL.}
\label{fig:y_delay_x_epochs}
\vspace*{-.5cm}
\end{figure}
Next, we analyze the convergence behavior of the proposed schemes, as shown in Fig.~\ref{fig:y_delay_x_epochs}, by setting $\sigma_h^2=0.008$, $\sigma_\omega^2=3200$, $B=3$ BSs, $L_b=6$ antennas per BS, and $K_b=5$ devices per BS over $500$ training epochs. The figure presents the convergence of the robust centralized learning scheme \emph{Robust Cen} and three curves representing the robust FL scheme \emph{Robust FL}, where each curve corresponds to one BS.
All schemes demonstrate convergence within the observed epochs. The key distinction lies in the comparison between \emph{Robust Cen} and \emph{Robust FL}. The centralized scheme, \emph{Robust Cen}, optimizes for centralized uncertainty and the overall worst-case delay across the network, leading to a higher final objective value (i.e., larger delay). In contrast, FL optimizes local objectives at individual BSs while estimating inter-BS interference, resulting in lower final objective values per BS, but not necessarily for the global robust worst-case delay $\hat{t}_\text{max}^{\hspace*{.02cm}\gamma}$.
Additionally, \emph{Robust FL} exhibits a sharp decline in the objective value between epochs $10$ and $40$, followed by another noticeable drop between epochs $200$ and $400$. This behavior suggests that Algorithm~\ref{alg:FLtrain} efficiently refines the model in early stages and benefits from further adjustments in later epochs.\vspace*{-.25cm}

\subsection{Impact of Uncertainty Samples}\label{ssec:param_N}\vspace*{-.0cm}
\begin{figure}[t]
\centering
\includegraphics[width=.9\linewidth]{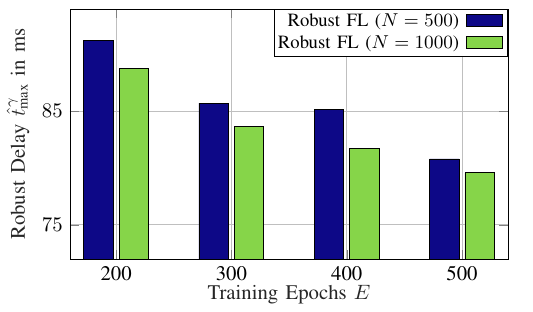}\\
\vspace*{-.25cm}
\caption{Robust worst-case delay $\hat{t}_\text{max}^{\hspace*{.02cm}\gamma}$ vs. training epochs showing the impact of uncertainty samples $N$.}
\label{fig:rev_y_delay_x_epochs_samplesN}
\vspace*{-.55cm}
\end{figure}
\begin{table}[t]
\vspace*{.2cm}
\renewcommand{\arraystretch}{1.2}
\centering
\begin{tabular}{r c c c c}
	\hline 
	\hspace*{-.2cm}\textbf{Training Epochs $\mathbf{E}$}\hspace*{-.2cm}& 
	{$\mathbf{200}$}\hspace*{-.2cm}& {$\mathbf{300}$}\hspace*{-.2cm}& 
	{$\mathbf{400}$}\hspace*{-.2cm}& 
	{$\mathbf{500}$} \\\hline
	$N=500$ \hspace*{-.2cm}& 
	4429 \hspace*{-.2cm}& 
	6657\hspace*{-.2cm}& 
	8885\hspace*{-.2cm}& 
	11110\hspace*{-.2cm}\\			
	$N=1000$ \hspace*{-.2cm}&
	8593 \hspace*{-.2cm}& 
	12906\hspace*{-.2cm}& 
	17222\hspace*{-.2cm}& 
	21533\hspace*{-.2cm}\\\hline
\end{tabular}
\vspace*{-.15cm}
\caption{Training time in seconds for \emph{Robust FL}.}
\label{tb:traintime}
\vspace{-.5cm}
\end{table}
{We investigate the impact of the number of uncertainty samples $N$ used in the joint uncertainty injection scheme. The simulation parameters are set to $\sigma_h^2 = 0.016$ and $\sigma_\omega^2 = 3200$. Fig.~\ref{fig:rev_y_delay_x_epochs_samplesN} shows the robust worst-case delay $\hat{t}_\text{max}^{\hspace*{.02cm}\gamma}$ as a function of the training epochs $E$ for different values of $N$. We consider $N \in \{500, 1000\}$, where $N=1000$ corresponds to the default configuration used throughout the paper. The results indicate that increasing $N$ improves both convergence stability and delay performance, as a larger number of samples provides a more accurate approximation of the underlying uncertainty distribution. In particular, choosing $N=1000$ consistently achieves a lower delay than $N=500$ across all epochs. However, the performance gap remains relatively small, suggesting that a moderate number of samples already yields a sufficiently accurate approximation of the robust objective. To further quantify the computational overhead, Tab.~\ref{tb:traintime} reports the corresponding training times. As expected, increasing the number of uncertainty samples from $N=500$ to $N=1000$ approximately doubles the training time, since the number of processed uncertainty realizations scales linearly with $N$. These findings highlight a trade-off between robustness and computational complexity, while also confirming that the proposed framework is not overly sensitive to the exact choice of $N$, i.e., it can operate efficiently with moderate sample sizes.}

\subsection{Training and Inference Data Exchange}\label{ssec:dat_ex}\vspace*{-.0cm}
\begin{figure}[t]
\centering
\includegraphics[width=.9\linewidth]{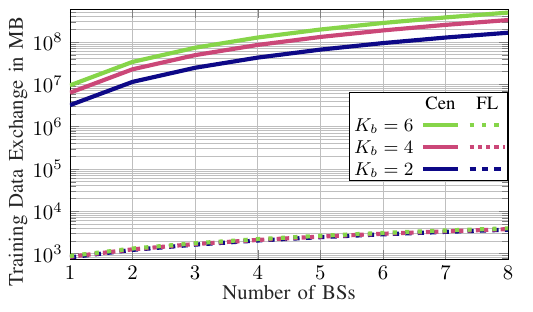}\\
\vspace*{-.3cm}
\caption{Exchanged data during training in MegaByte (MB) versus the number of BSs for different device numbers.}
\label{fig:y_logdatsize_x_numBS_training}
\vspace*{-.55cm}
\end{figure}
\begin{figure}[t]
\centering
\includegraphics[width=.9\linewidth]{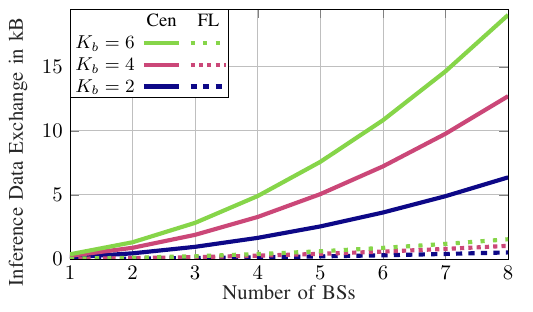}\\
\vspace*{-.3cm}
\caption{Exchanged data for inference in kiloByte (kB) versus the number of BSs for different device numbers.}
\label{fig:y_logdatsize_x_numBS_inference}
\vspace*{-.55cm}
\end{figure}
The proposed centralized and FL-based schemes rely on the exchange of data through limited-capacity backhaul network, which connects BSs and CP. {The total data exchange required for ML training is depicted in Fig.~\ref{fig:y_logdatsize_x_numBS_training}, as a function of the number of BSs for different values of devices per BS. The y-axis of Fig.~\ref{fig:y_logdatsize_x_numBS_training} represents the training data exchange (in megabytes (MB)) for both centralized (i.e., Algorithm~\ref{alg:CENtrain}) and federated (i.e., Algorithm~\ref{alg:FLtrain}) implementations.}
{While the data exchange required for training Algorithm~\ref{alg:FLtrain} remains within the range of $800$-$4000$ MB, the centralized approach requires up to $495000000$ MB (i.e., $495$ TB) when the network consists of $8$ BSs with $K_b=6$.} Furthermore, the required data exchange in the centralized implementation grows significantly faster than in the FL-based schemes. Algorithm~\ref{alg:CENtrain} does not require exchanging the DNN model with the CP but necessitates frequent transmissions of local channel and computation estimations for each training/testing realization. Algorithm~\ref{alg:FLtrain} only transmits DNN model parameters and inter-BS interference levels, keeping excessive estimations local. Moreover, Fig.~\ref{fig:y_logdatsize_x_numBS_training} highlights the impact of increasing $K_b$, which further amplifies data exchange requirements in both schemes. 

The limited-capacity BS-CP backhaul network may become a major bottleneck in FL training. This overhead can be reduced through communication-efficient FL techniques such as probabilistic device selection, parameter quantization, model compression, or sparse/partial model updates \cite{doi:10.1073/pnas.2024789118, kim2024spafl}. Such approaches reduce the number of transmitted bits per aggregation round, at the cost of introducing approximation errors into the local or aggregated model updates. Hence, overly aggressive compression may slow convergence or degrade the learned resource-allocation policy, potentially increasing the resulting robust delay. Investigating this communication-robustness trade-off is an interesting direction for future work.

In practical implementations, the training phase constitutes only a fraction of the overall operational data exchange. 
Fig.~\ref{fig:y_logdatsize_x_numBS_inference} illustrates the inference-phase data exchange requirements (in kilobytes (kB)) versus the number of BSs for different values of $K_b$. Both centralized and FL-based schemes initially exhibit low inference data exchange requirements. However, as the number of BSs increases, the centralized scheme's data exchange grows exponentially, while the federated data exchange exhibits a much slower increase. Specifically, at $B=8$ and $K_b=6$, Fig.~\ref{fig:y_logdatsize_x_numBS_inference} shows that the centralized approach requires $19$ kB, whereas the distributed one requires only $1.5$ kB. Additionally, the impact of higher $K_b$ is more pronounced in the centralized scheme. 
These findings emphasize that the proposed FL-based robust delay minimization approach requires significantly less data exchange in both training and inference phases, making it a more practical and scalable solution for future data-driven, distributed XR systems.\vspace*{-.2cm}

\subsection{Ablation Study}
\begin{figure}[t]
\centering
\includegraphics[width=.99\linewidth]{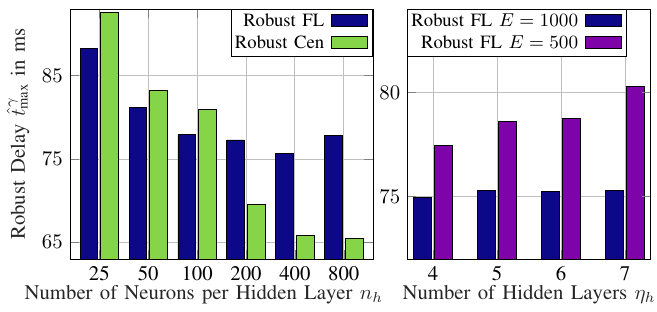}\\
\vspace*{-.25cm}
\caption{Robust worst-case delay $\hat{t}_\text{max}^{\hspace*{.02cm}\gamma}$ vs. neurons per hidden layer $n_h$ (left) and number of hidden layers $\eta_h$ (right).}
\label{fig:rev_y_delay_x_ablation}
\vspace*{-.55cm}
\end{figure}
{As a last simulation set, we investigate the impact of the DNN architecture on the achieved robustness performance. The simulation parameters are set to $\sigma_h^2 = 0.016$ and $\sigma_\omega^2 = 3200$. Fig.~\ref{fig:rev_y_delay_x_ablation} illustrates the robust worst-case delay $\hat{t}_\text{max}^{\hspace*{.02cm}\gamma}$ as a function of the number of neurons per hidden layer $n_h$ (left) and the number of hidden layers $\eta_h$ (right).
For the left-hand side of Fig.~\ref{fig:rev_y_delay_x_ablation}, the number of hidden layers is fixed to $\eta_h=10$ for \emph{Robust Cen} and $\eta_h=7$ for \emph{Robust FL}. The results of Fig.~\ref{fig:rev_y_delay_x_ablation} show that increasing the number of neurons generally improves the delay performance for both schemes. However, saturation effects can be observed for larger network sizes, where increasing $n_h$ beyond a certain point provides only marginal gains. Furthermore, \emph{Robust Cen} benefits more strongly from larger DNN sizes, reflecting the increased complexity of the centralized optimization task.
For the right-hand side of Fig.~\ref{fig:rev_y_delay_x_ablation}, the number of neurons per hidden layer is fixed to $n_h=200$, while the number of hidden layers $\eta_h$ is varied for the \emph{Robust FL} scheme. The results show that deeper architectures may negatively affect performance when the training duration is insufficient, whereas for longer training durations, the impact of $\eta_h$ becomes less significant. Overall, the results highlight a trade-off between model complexity, training effort, and achieved robustness performance.}

\subsection{Edge Deployment Feasibility}
To further assess the practical deployment feasibility of the proposed DNN-based resource-allocation schemes, we evaluate their inference latency on edge hardware. The measurements are conducted on an NVIDIA Jetson Orin Nano Developer Kit Super operated in 25~W power mode, running JetPack~7.2.1, CUDA~13.2, Python~3.12.3, and PyTorch~2.12.0+cu132. The platform is equipped with a 1024-core NVIDIA Ampere GPU, 32 Tensor Cores, and 8~GB of unified LPDDR5 memory.	Inference is performed with batch size ``one" to reflect online resource allocation for a single instantaneous network state. After 50 warm-up runs, the average inference latency of \emph{Robust Cen} is $9.578$~ms with a standard deviation of $0.12$~ms, while \emph{Robust FL} achieves an average inference latency of $4.587$~ms with a standard deviation of $0.06$~ms. These results demonstrate that both architectures support millisecond-scale inference on resource-constrained edge hardware. In particular, the lower latency of \emph{Robust FL} reflects the reduced input dimension and smaller local DNN architecture compared with the centralized scheme. Hence, once trained offline, the proposed models can be deployed for online resource-allocation decisions without requiring computationally expensive optimization during operation.

\section{Conclusion}\label{sec:con}\vspace*{-.1cm}
Automating the decision-making via FL and distributed intelligence emerges as a key enabler toward seamless future XR-based networks. This paper presents a novel deep learning-based approach to jointly optimize the communication and computation resources in both centralized and federated settings while accounting for uncertainties in channel estimation and computing requirements. The proposed robust framework manages transmit power, computing power, and computation allocation through a DNN trained with uncertainty injection. By jointly incorporating uncertainty samples during training and optimizing resource allocations based on the robust utility, our approach effectively minimizes the worst-case delay across multiple devices. Our findings highlight the benefits of joint uncertainty injection, demonstrating that FL-based solutions achieve near-centralized performance while significantly enhancing scalability, efficiency, and robustness. These results particularly highlight the potential of AI-driven, multi-purpose, distributed resource management for real-world XR networks. The results of the current paper further trigger a handful of future research directions in the field, e.g., optimizing other objectives of practical interest within the joint communication and computation interplay{, considering hybrid and online learning protocols,} addressing different system models of emerging interest within the context of non-terrestrial networks{, the impact of stale interference information} (among others). \vspace*{-.5cm}

\appendix
\section{Benchmark Optimization Problems}
For the sake of completeness, this appendix presents a handful of data-driven as well as numerical optimization baselines, which are used as benchmarking schemes in the simulations section of the paper. The methods, hereafter called \emph{Non-robust Cen}, \emph{Non-robust FL}, \emph{Non-robust FP}, \emph{Uniform FP}, \emph{SAA/WMMSE}, {\emph{FedProx}, and \emph{Local}}, rely on specific formulations of the optimization problem, as discussed next.\vspace*{-.3cm}

\subsection{Non-Robust Centralized Learning}\label{appendix_a}\vspace*{-.1cm}
The non-robust counterpart of problem (11) rather neglects the channel and computing uncertainties in (1) and (5), and minimizes the nominal worst-case delay $t_\text{max}$ among all devices:\vspace*{-.3cm}
\begin{subequations}\label{eq:Opt1_reg}
	\begingroup
	\addtolength{\jot}{-.1cm}
	\begin{align}
		\underset{\bm{p}^\text{tx},\bm{p}^\text{co},\bm{f}^\text{co}}{\text{min}}\quad & \underset{k\in\mathcal{K}}{\text{max}}\{t_k\} \tag{\ref{eq:Opt1_reg}} \\
		\text{s.t.} \quad\quad & (6),(9), \hspace*{-.2cm} \nonumber\\[-.65cm]\nonumber
	\end{align}
	\endgroup
\end{subequations}
with computation capacity (6) and power constraint (9). Unlike the robust formulation (11), the $\gamma$-th quantile constraint (10) is omitted here.
The \emph{Non-robust Cen} benchmark scheme can be derived from problem \eqref{eq:Opt1_reg} by applying Algorithm~1 without uncertainty injection, i.e., by replacing $\hat{t}_\text{max}^{\hspace*{.02cm}\gamma}$ with $\underset{k\in\mathcal{K}}{\text{max}}\{t_k\}$.\vspace*{-.2cm}

\subsection{Non-Robust Federated Learning}\vspace*{-.1cm}
The non-robust version of the distributed problem (18) neglects channel and computing uncertainties, and minimizes the nominal worst-case delay $t_{b,\text{max}}$ among all devices associated with BS $b$:\vspace*{-.4cm}
\begin{subequations}\label{eq:Opt2_reg}
	\begingroup
	\addtolength{\jot}{-.1cm}
	\begin{align}
		\underset{\bm{p}^\text{tx}_b,p^\text{co}_b,\bm{f}^\text{co}_b}{\text{min}}\quad & \underset{k\in\mathcal{K}_b}{\text{max}}\{t_{k}\} \tag{\ref{eq:Opt2_reg}} \\
		\text{s.t.} \quad\quad & (6),(9), (18b) \hspace*{-.2cm} \nonumber\\[-.65cm]\nonumber
	\end{align}
	\endgroup
\end{subequations}
with computation capacity (6) constraint, power constraint (9), and inter-BS interference constraint (18b). The \emph{Non-robust FL} benchmark scheme can be derived from problem \eqref{eq:Opt2_reg} by applying Algorithm~2 without uncertainty injection.

\subsection{Non-Robust Numerical Optimization}\vspace*{-.1cm}
To compare the data-driven ML-based schemes to a numerical optimization approach, \emph{Non-robust FP} utilizes FP \cite{8314727} to solve problem \eqref{eq:Opt1_reg} using standard solvers (e.g., CVX \cite{cvx}).\vspace*{-.2cm}

\subsection{Uniform Numerical Optimization}\vspace*{-.1cm}
A modified optimization formulation based on problem \eqref{eq:Opt1_reg} with additional constraints enforcing equal transmit power and computation cycles is formulated as:\vspace*{-.2cm}
\begin{subequations}\label{eq:Opt1_uni}
	\begingroup
	\addtolength{\jot}{-.1cm}
	\begin{align}
		\underset{\bm{p}^\text{tx},\bm{p}^\text{co},\bm{f}^\text{co}}{\text{min}}\quad & \underset{k\in\mathcal{K}}{\text{max}}\{t_k\} \tag{\ref{eq:Opt1_uni}} \\
		\text{s.t.} \quad\quad & (6),(9), \hspace*{-.2cm} \nonumber\\
		&p_k^\text{tx} = p_{k'}^\text{tx},\, f_k^\text{co} = f_{k'}^\text{co},\quad \forall k\neq k'\in\mathcal{K}.\label{eq:uni}\\[-.65cm]\nonumber
	\end{align}
	\endgroup
\end{subequations}
Constraint \eqref{eq:uni} enforces all devices to be served with the same transmit power $p_k^\text{tx}$ and computing allocation $f_k^\text{co}$, $\forall k\in\mathcal{K}$. \emph{Uniform FP} then solves \eqref{eq:Opt1_uni} using FP \cite{8314727} optimizing solely the power split between transmission and computation.\vspace*{-.2cm}

\subsection{Robust Numerical Optimization}\label{appendix_b}\vspace*{-.1cm}
Finally, for the \emph{SAA/WMMSE} scheme, similar to the uncertainty injection scheme proposed in Sec.~III-B, $N$ realizations of channel and computing errors $\bm{h}^\text{err}_n$ and $\bm{\omega}^\text{err}_n$, $n\in\{1,\dots,N\}$, are sampled, yielding $N$ worst-case delays $t_{\text{max},n}$ as in (16). The resulting SAA-based problem is then formulated as follows:\vspace*{-.2cm}
\begin{subequations}\label{eq:Opt3}
	\begingroup
	\addtolength{\jot}{-.1cm}
	\begin{align}
		\underset{\bm{p}^\text{tx},\bm{p}^\text{co},\bm{f}^\text{co}}{\text{min}}\quad &t_\text{max}^\text{SAA} \tag{\ref{eq:Opt3}} \\
		\text{s.t.} \quad\quad & (1),(5),(6),(9),\hspace*{-.2cm} \nonumber\\
		&t_\text{max}^\text{SAA} = \frac{1}{N} \sideset{}{_{n=1}^{N}}{\sum} t_{\text{max},n}, &\label{eq:t_SAA}\\[-.65cm]\nonumber
	\end{align}
	\endgroup
\end{subequations}
where the $\gamma$-th quantile constraint (10) is replaced by the SAA constraint \eqref{eq:t_SAA}. Following \cite{reifert2021rate}, problem \eqref{eq:Opt3} can then be solved via an SAA and WMMSE-based method.

\subsection{Robust FedProx-based Learning}
{The work \cite{MLSYS2020} proposes the Federated Proximal (FedProx) algorithm to address optimization in heterogeneous networks. The key idea is that, in each communication round, all agents receive the global model, use it to initialize their local models, and perform local training with an additional proximal penalty term that discourages large deviations from the global model.}

{To adopt the FedProx algorithm for solving problem (18), we modify the procedure of Algorithm~2 as follows. In particular, the local model updates in Step~$3$ of Algorithm~2 are augmented by a modified gradient $\partial\mathcal{L}_b/\partial\bm{\Theta}_b$, based on the following loss function:
	\begin{equation}\label{eq:lossFedProx}
		\mathcal{L}_b = \hat{t}_{b,\text{max}}^{\hspace*{.02cm}\gamma} + \frac{\delta}{2} \lVert\bm{\Theta}_b-\bm{\Theta}^\text{global}\rVert^2.
	\end{equation}	 
	Here, $\delta$ is a regularization parameter that controls how strongly the local model is pulled toward the global model, i.e., it penalizes local model drift. During inference, all BSs use the final aggregated global model, ensuring consistent and generalizable performance across the network. In our implementation, we set $\delta=10^{-3}$, which provides a trade-off between the standard FedAvg behavior (small $\delta$) and a strongly constrained local update (large $\delta$).}

{Extensions toward heterogeneous model architectures, e.g., HeteroFL \cite{diao2020heterofl}, may further improve scalability by adapting model complexity to BS capabilities, but are beyond the scope of this work.}\vspace*{-.3cm}

\subsection{Fully Local Learning}
{As an additional benchmark, particularly for heterogeneous network scenarios, we consider a fully local learning approach. The key idea is to omit the global model aggregation step and maintain separate models at each BS. The \emph{Local} learning scheme is obtained as a modified version of Algorithm~2, where the global aggregation in Step~7 is omitted. Consequently, Steps~5 and~9 of Algorithm~2 are performed without the use of a global model $\bm{\Theta}^\text{global}$, i.e., each BS updates and maintains its own local model independently. Note that this local learning scheme still relies on computing, collecting, and exchanging inter-BS interference information.}\vspace*{-.4cm}

\bibliographystyle{IEEEtran}
\bibliography{bibliography}

\end{document}